\documentclass[prb,tighten]{revtex4}

\usepackage{amssymb,amsmath}

\usepackage{epsfig}

\usepackage{graphicx}

\begin{document}
\title{Navier-Stokes transport coefficients of $d$-dimensional granular binary mixtures
at low density}
\author{Vicente Garz\'{o}\footnote[1]{Electronic address: vicenteg@unex.es;
URL: http://www.unex.es/eweb/fisteor/vicente/}}
\affiliation{Departamento de F\'{\i}sica, Universidad de
Extremadura, E-06071 Badajoz, Spain}
\author{Jos\'e Mar\'{\i}a
Montanero\footnote[2] {Electronic address: jmm@unex.es}}
\affiliation{Departamento de Electr\'onica e Ingenier\'{\i}a
Electromec\'anica, Universidad de Extremadura, E-06071 Badajoz,
Spain}

\begin{abstract}
The Navier-Stokes transport coefficients for binary mixtures of smooth inelastic hard disks or spheres under
gravity are determined from the Boltzmann kinetic theory by application of the Chapman-Enskog method for states
near the local homogeneous cooling state. It is shown that the Navier-Stokes transport coefficients are not
affected by the presence of gravity. As in the elastic case, the transport coefficients of the mixture verify a
set of coupled linear integral equations that are approximately solved by using the leading terms in a Sonine
polynomial expansion. The results reported here extend previous calculations [V. Garz\'o and J. W. Dufty, Phys.
Fluids {\bf 14}:1476--1490 (2002)] to an arbitrary number of dimensions and provide explicit expressions for the
seven Navier-Stokes transport coefficients in terms of the coefficients of restitution and the masses,
composition, and sizes of the constituents of the mixture. In addition, to check the accuracy of our theory, the
inelastic Boltzmann equation is also numerically solved by means of the direct simulation Monte Carlo method to
evaluate the diffusion and shear viscosity coefficients for hard disks. The comparison shows a good agreement
over a wide range of values of the coefficients of restitution and the parameters of the mixture (masses and
sizes).


 Running title: Navier-Stokes transport coefficients

\end{abstract}

\draft
\date{\today}
\maketitle

\section{Introduction}
\label{sec1}

The simplest model for a granular fluid is a system composed by
smooth hard spheres or disks with inelastic collisions. The only
difference from the corresponding model for normal fluids is the
loss of energy in each binary collision, characterized by a
(constant) coefficient of normal restitution. For a low density
gas, the Boltzmann kinetic equation conveniently modified to
account for inelastic collisions \cite{GS95,BDS97,BP04} has been
used in recent years as the starting point to derive the
hydrodynamic-like equations of the system. Thus, assuming the
existence of a {\em normal} (hydrodynamic) solution for
sufficiently long space and time scales, the Chapman-Enskog (CE)
method \cite{CC70} has been applied to solve the Boltzmann
equation to Navier-Stokes (NS) order and get explicit expressions
for the transport coefficients. While this goal has been widely
covered in the case of a monocomponent gas, \cite{simple} much
less is known for systems composed by grains of different masses,
sizes, and concentrations (granular mixtures).

Needless to say, the determination of the NS transport
coefficients for a multicomponent granular fluid is much more
complicated than for a single granular system. Many attempts to
derive these coefficients \cite{mixture} have been carried out by
means of the CE expansion around Maxwellians at the {\em same}
temperature $T$ for each species. The use of this distribution can
only be considered as acceptable for nearly elastic systems where
the assumption of the equipartition of energy still holds. In
addition, according to this level of approximation, the
inelasticity is only accounted for by the presence of a sink term
in the energy balance equation and so the expressions of the NS
transport coefficients are the same as those obtained for elastic
collisions. However, as the dissipation increases, different
species of a granular mixture have different partial temperatures
$T_i$ and consequently, the energy equipartition is seriously
broken ($T_i\neq T$). The failure of energy equipartition in
granular fluids \cite{GD99b,MP99} has also been confirmed by
computer simulations \cite{computer} and even observed in real
experiments \cite{exp1} of agitated mixtures. All the results show
that deviations from equipartition depend on the mechanical
differences between the particles of each species and the
coefficients of restitution of the system. Given that the
inclusion of nonequipartition effects increases the level of
complexity of the problem, it is interesting from a practical
point of view to assess the influence of this effect on the
transport properties of the system. If the NS transport
coefficients turned out to be quite sensitive to nonequipartition,
the predictions made from previous theories \cite{mixture} should
be reexamined by theories that take into account the
nonequipartition of energy. For this reason, although the
possibility of nonequipartition was already pointed out many years
ago, \cite{JM87} a careful study of its influence on transport has
only been carried out recently. In this context, Garz\'o and Dufty
\cite{GD02} have developed a kinetic theory for a binary granular
mixture of inelastic hard spheres at low density which accounts
for nonequipartition effects. Their results show that in general
the consequences of the temperature differences for the transport
coefficients are quite significant, especially for strong
dissipation.\cite{GA05,MGD06} It is important to remark that the
expressions derived in Ref.\ \onlinecite{GD02} for the
Navier-Stokes transport coefficients do not limit their
application to weak inelasticity. In fact, the results reported in
this paper include a domain of both weak and strong inelasticity,
$0.5 \leq \alpha \leq 1$, where $\alpha$ is the (common)
coefficient of restitution considered. The accuracy of these
theoretical predictions (based on a Sonine polynomial expansion)
has been confirmed by Monte Carlo simulations of the inelastic
Boltzmann equation in the cases of the diffusion coefficient
\cite{GM04} and the shear viscosity coefficient of a mixture
heated by an external thermostat. \cite{MG03} Exceptions to this
good agreement are extreme mass or size ratios and strong
dissipation, although these discrepancies are mitigated in part if
one retains more terms in the Sonine polynomial expansion.
\cite{GM04} For small dissipation, the results derived by Garz\'o
and Dufty \cite{GD02} agree with those recently obtained by Serero
{\em et al.}\cite{SGNT06} in the first order of the order
parameter $\epsilon_{ij}=1-\alpha_{ij}^2$.

The CE method solves the Boltzmann equation by expanding the
distribution function of each species $f_i({\bf r}, {\bf v},t)$
around the {\em local} homogeneous cooling state (HCS).
\cite{GD99b} This state plays the same role for granular gases as
the local equilibrium distribution for a gas with elastic
collisions. Given that the form of the distribution function
$f_i^{(0)}$ of the HCS is not exactly known, one usually considers
the first correction to a Maxwellian at the temperature for that
species, namely, a polynomial in velocity of degree four (leading
Sonine correction). However, the results derived for hard spheres
clearly show that the influence of these non-Gaussian
contributions to the transport coefficients are in general
negligible, except in the case of the heat flux for quite large
values of dissipation. \cite{MGD06} Accordingly, a theory
incorporating the contributions coming from the deviations of the
HCS from its Gaussian form does not seem necessary in practice for
computing the NS transport coefficients of the mixture.

The objective of this paper is twofold. First, given that the
results reported in Ref.\ \onlinecite{GD02} are limited to hard
spheres, we extend here this derivation to an arbitrary number of
dimensions $d$. This goal is not only academic since, from a
practical standpoint, many of the experiments reported for flowing
granular materials have created (quasi) {\em two-dimensional}
systems by confining grains between vertical or on a horizontal or
tilted surface, enabling data collection by high-speed video.
\cite{exp} Regarding computer simulations, most of them consider
hard disks to save computer time and memory. For these reasons, it
would be desirable to provide experimentalists and simulators with
theoretical tools to work when studying problems both in two and
three dimensions. In addition, apart from its practical interest,
it is also interesting from a fundamental view to explore what is
the influence of dimensionality on the dependence of the transport
coefficients on dissipation. As a second target, we want also to
present a simplified theory with explicit expressions for the
transport coefficients. As the algebra involved in the
calculations of Ref.\ \onlinecite{GD02} is complex, the
constitutive relations for the fluxes were not explicitly
displayed in this paper. Although the work carried out here
involves complex algebra as well, the use of Maxwellians at
different temperatures for the distribution functions of each
species in the reference state allows us to explicitly obtain
expressions for the {\em seven} relevant transport coefficients of
the mixture in terms of the mechanical parameters of the system:
masses, sizes, composition and coefficients of restitution. To
assess the degree of accuracy of our (approximated) expressions,
we have also performed Monte Carlo simulations for the diffusion
and the shear viscosity coefficients for hard disks ($d=2$). As
shown below, the good agreement found between the results derived
in this paper with computer simulations justifies this
simplification and allows one to obtain more simplified forms of
the transport coefficients.

The plan of the paper is as follows. In Sec.\ \ref{sec2}, the
inelastic Boltzmann equation and the corresponding hydrodynamic
equations are recalled. The CE expansion adapted to the inelastic
binary mixtures is formulated in Sec.\ \ref{sec2bis}. Assuming
that gradients and dissipation are independent parameters, the
Boltzmann equation is solved by means of an expansion in powers of
the spatial gradients around the local HCS distribution
$f_i^{(0)}$. It is shown that the use of the local HCS as the
reference state is not an assumption of the CE method but a
consequence of the exact solution to the Boltzmann equation in the
zeroth-order approximation. Section \ref{sec3} deals with the
expressions for the NS transport coefficients. As in the case of
elastic collisions, these coefficients are the solutions of a set
of coupled linear integral equations which involves the (unknown)
distributions $f_i^{(0)}$. The integral equations are solved by
considering two approximations: First, $f_i^{(0)}$ is replaced by
its Maxwellian form at the temperature $T_i$, and second, only the
leading terms in a Sonine polynomial expansion of the first-order
distribution $f_i^{(1)}$ are retained. Technical details of the
calculations carried out here are given in Appendices \ref{appA},
\ref{appB}, and \ref{appC}. A comparison with previous results
\cite{SGNT06} based on the use of Maxwellians at the same
temperature $T$ as a ground state is also illustrated in Sec.\
\ref{sec3}, showing significant discrepancies between both
descriptions at moderate dissipation. Section \ref{sec4} is
devoted to the numerical solutions of the Boltzmann equation by
using the direct simulation Monte Carlo (DSMC) method \cite{Bird}
in the cases of the diffusion $D$ and shear viscosity $\eta$
coefficients for hard disks. To the best of our knowledge, this is
the first time that the NS shear viscosity of a granular binary
mixture at low density has been numerically obtained from the DSMC
method. The paper is closed in Sec.\ \ref{sec5} with a brief discussion of the
results presented in this paper.

\section{Boltzmann equation and conservation laws}
\label{sec2}

Consider a binary mixture composed by smooth inelastic disks
($d=2$) or spheres ($d=3$) of masses $m_{1}$ and $ m_{2} $, and
diameters $\sigma _{1}$ and $\sigma _{2}$. The inelasticity of
collisions among all pairs is characterized by three independent
constant coefficients of restitution $\alpha _{11}$, $\alpha
_{22}$, and $\alpha _{12}=\alpha _{21}$, where $\alpha _{ij}\leq
1$ is the coefficient of restitution for collisions between
particles of species $i$ and $j$. The mixture is in presence of
the gravitational field so that each particle feels the action of
the force ${\bf F}_i=m_i{\bf g}$, where ${\bf g}$ is the gravity
acceleration. In the low density regime, the distribution
functions $f_{i}({\bf r},{\bf v};t)$ $(i=1,2)$ for the two species
are determined from the set of nonlinear Boltzmann equations
\cite{BDS97}
\begin{equation}
\left( \partial _{t}+{\bf v}\cdot\nabla+{\bf g} \cdot
\frac{\partial}{\partial {\bf v}} \right) f_{i}({\bf r},{\bf
v},t)=\sum_{j=1}^2 J_{ij}\left[ {\bf v}|f_{i}(t),f_{j}(t)\right]
\;, \label{2.1}
\end{equation}
where the Boltzmann collision operator $J_{ij}\left[ {\bf
v}|f_{i},f_{j}\right] $ is
\begin{eqnarray}
J_{ij}\left[ {\bf v}_{1}|f_{i},f_{j}\right] &=&\sigma
_{ij}^{d-1}\int d{\bf v} _{2}\int d\widehat{\boldsymbol {\sigma
}}\,\Theta (\widehat{{\boldsymbol {\sigma }}} \cdot {\bf
g}_{12})(\widehat{\boldsymbol {\sigma }}\cdot {\bf g}_{12})
\nonumber
\\
&&\times \left[ \alpha _{ij}^{-2}f_{i}({\bf r},{\bf v}_{1}^{\prime
},t)f_{j}( {\bf r},{\bf v}_{2}^{\prime },t)-f_{i}({\bf r},{\bf v}
_{1},t)f_{j}({\bf r}, {\bf v}_{2},t)\right] \;. \label{2.2}
\end{eqnarray}
In Eq.\ (\ref{2.2}), $d$ is the dimensionality of the system,
$\sigma _{ij}=\left( \sigma _{i}+\sigma _{j}\right) /2$,
$\widehat{\boldsymbol {\sigma}}$ is an unit vector along the line
of centers, $\Theta $ is the Heaviside step function, and ${\bf
g}_{12}={\bf v}_{1}-{\bf v}_{2}$ is the relative velocity. The
primes on the velocities denote the initial values $\{{\bf
v}_{1}^{\prime }, {\bf v}_{2}^{\prime }\}$ that lead to $\{{\bf
v}_{1},{\bf v}_{2}\}$ following a binary (restituting) collision:
\begin{equation}
{\bf v}_{1}^{\prime }={\bf v}_{1}-\mu _{ji}\left( 1+\alpha
_{ij}^{-1}\right) (\widehat{{\boldsymbol {\sigma }}}\cdot {\bf
g}_{12})\widehat{{\boldsymbol {\sigma }}} ,\nonumber\\
\end{equation}
\begin{equation}
 {\bf v}_{2}^{\prime }={\bf v}_{2}+\mu _{ij}\left( 1+\alpha
_{ij}^{-1}\right) (\widehat{{\boldsymbol {\sigma }}}\cdot {\bf
g}_{12})\widehat{ \boldsymbol {\sigma}} ,  \label{2.3}
\end{equation}
where $\mu _{ij}\equiv m_{i}/\left( m_{i}+m_{j}\right) $. The
relevant hydrodynamic fields are the number densities $n_{i}$, the
flow velocity $ {\bf u}$, and the temperature $T$. They are
defined in terms of moments of the distributions $f_{i}$ as
\begin{equation}
n_{i}=\int d{\bf v}f_{i}({\bf v})\;,\quad \rho {\bf u}=\sum_{i=1}^2m_{i}
\int d {\bf v}{\bf v}f_{i}({\bf v})\;,
\label{2.4}
\end{equation}
\begin{equation}
nT=p=\sum_{i=1}^2 n_i T_i=\sum_{i=1}^2\frac{m_{i}}{d}\int d{\bf
v}V^{2}f_{i}({\bf v})\;, \label{2.5}
\end{equation}
where ${\bf V}={\bf v}-{\bf u}$ is the peculiar velocity, $
n=n_{1}+n_{2}$ is the total number density, $\rho
=m_{1}n_{1}+m_{2}n_{2}$ is the total mass density, and $p$ is the
pressure. Furthermore, the third equality of Eq.\ (\ref{2.5})
defines the kinetic temperatures $T_i$ for each species, which
measure their mean kinetic energies.

The collision operators conserve the particle number of each
species and the total momentum but the total energy is not
conserved:
\begin{equation}
\int d{\bf v}J_{ij}[{\bf v}|f_{i},f_{j}]=0\;,  \label{2.6}
\end{equation}
\begin{equation}
\sum_{i=1}^2\sum_{j=1}^2m_i\int d{\bf v}{\bf v}J_{ij}[{\bf
v}|f_{i},f_{j}]=0 \;, \label{2.7}
\end{equation}
\begin{equation}
\sum_{i=1}^2\sum_{j=1}^2m_i\int d{\bf v}V^{2}J_{ij}[{\bf v}
|f_{i},f_{j}]=-d nT\zeta \;,  \label{2.8}
\end{equation}
where $\zeta$ is identified as the total ``cooling rate'' due to
inelastic collisions among all species. At a kinetic level, it is
also convenient to introduce the ``cooling rates'' $\zeta_i$ for
the partial temperatures $T_i$. They are defined as
\begin{equation}
\label{2.7.1} \zeta_i=\sum_{j=1}^2\zeta_{ij}=-
\frac{m_i}{dn_iT_i}\sum_{j=1}^2\int d{\bf v}V^{2}J_{ij}[{\bf
v}|f_{i},f_{j}],
\end{equation}
where the second equality defines the quantities $\zeta_{ij}$. The
total cooling rate $\zeta$ can be written in terms of the partial
cooling rates $\zeta_i$ as
\begin{equation}
\label{2.7.2} \zeta=T^{-1}\sum_{i=1}^2\;x_iT_i\zeta_i,
\end{equation}
where $x_i=n_i/n$ is the mole fraction of species $i$.

From Eqs.\ (\ref{2.4})--(\ref{2.8}), the macroscopic balance
equations for the mixture can be obtained. They are given by
\begin{equation}
D_{t}n_{i}+n_{i}\nabla \cdot {\bf u}+\frac{\nabla \cdot {\bf
j}_{i}}{m_{i}} =0\;,  \label{2.9}
\end{equation}
\begin{equation}
D_{t}{\bf u}+\rho ^{-1}\nabla \cdot {\sf P}={\bf g}\;,
\label{2.10}
\end{equation}
\begin{equation}
D_{t}T-\frac{T}{n}\sum_{i=1}^2\frac{\nabla \cdot {\bf
j}_{i}}{m_{i}}+\frac{2}{dn} \left( \nabla \cdot {\bf q}+{\sf
P}:\nabla {\bf u}\right) =-\zeta \,T\;. \label{2.11}
\end{equation}
In the above equations, $D_{t}=\partial _{t}+{\bf u}\cdot \nabla $
is the material derivative,
\begin{equation}
{\bf j}_{i}=m_{i}\int d{\bf v}\,{\bf V}\,f_{i}({\bf v})
\label{2.11b}
\end{equation}
is the mass flux for species $i$ relative to the local flow,
\begin{equation}
{\sf P}=\sum_{i=1}^2\,m_i\,\int d{\bf v}\,{\bf V}{\bf
V}\,f_{i}({\bf v})  \label{2.12}
\end{equation}
is the total pressure tensor, and
\begin{equation}
{\bf q}=\sum_{i=1}^2\,\frac{m_i}{2}\int d{\bf v}\,V^{2}{\bf
V}\,f_{i}({\bf v})  \label{2.13}
\end{equation}
is the total heat flux.

The macroscopic balance equations (\ref{2.9})--(\ref{2.11}) are
not entirely expressed in terms of the hydrodynamic fields, due to
the presence of the cooling rate $\zeta$, the mass flux ${\bf
j}_i$, the heat flux ${\bf q}$, and the pressure tensor ${\sf P}$
which are given as functionals of the distributions $f_i$.
However, it these distributions can be expressed as functionals of
the hydrodynamic fields, then the cooling rate and the fluxes also
will become functional of the hydrodynamic fields through Eqs.\
(\ref{2.7.1}) and (\ref{2.11b})--(\ref{2.13}). Such expressions
are called {\em constitutive} relations and they provide a link
between the exact balance equations and a closed set of equations
for the hydrodynamic fields. This hydrodynamic description can be
derived by looking for a {\em normal} solution to the Boltzmann
kinetic equation. A normal solution is one whose all space and
time dependence of the distribution function $f_i$ occurs through
a functional dependence on the hydrodynamic fields,
\begin{equation}
f_{i}({\bf r},{\bf v},t)=f_{i}\left[{\bf v}|x_{1} ({\bf r}, t),
p({\bf r}, t), T({\bf r}, t), {\bf u}({\bf r}, t) \right] \;.
\label{2.14}
\end{equation}
As in previous works, \cite{GD02,MGD06} we have taken the set
$\{x_1, p, T, {\bf u}\}$ as the $d+3$ independent fields of the
two-component mixture. These are the most accessible fields from
an experimental point of view. The determination of this normal
solution from the Boltzmann equation (\ref{2.1}) is a very
difficult task in general, unless the spatial gradients are small.
In this case, the CE method gives an approximate solution.

\section{Chapman-Enskog solution}
\label{sec2bis}

The CE method is a procedure to construct an approximate normal
solution. It is perturbative, using the spatial gradients as the
small expansion parameter. More specifically, one assumes that the
spatial variations of the hydrodynamic fields $n_i$, ${\bf u}$,
$p$, and $T$ are small on the scale of the mean free path. For
ordinary gases this can be controlled by the initial or boundary
conditions. It is more complicated for granular gases, since in
some cases (e.g., steady states such as the simple shear flow
problem\cite{SGD04}) the boundary conditions imply a relationship
between dissipation and some hydrodynamic gradient. As a
consequence, there are examples for which the NS approximation is
restricted to the quasi-elastic limit.\cite{SGD04} Here, we also
assume that the spatial gradients are independent of the
coefficients of restitution so that, the corresponding NS order
hydrodynamic equations apply for small gradients but they are not
limited {\em a priori} to weak inelasticity. It must be emphasized
that our perturbation scheme differs from the one recently carried
out by Serero {\em et al.} \cite{SGNT06} where the CE solution is
given in powers of both the hydrodynamic gradients (or
equivalently, the Knudsen number) and the degree of dissipation
$\epsilon_{ij}=1-\alpha_{ij}^2$. The results provided in Ref.\
\onlinecite{SGNT06} only agree with our results in the
quasielastic domain (small $\epsilon_{ij}$). Moreover, in the
presence of an external force it is necessary to characterize the
magnitude of the force relative to gradients as well. As in the
elastic case, \cite{CC70} it is assumed here that the magnitude of
the gravity field is at least of first order in perturbation
expansion.

For small spatial variations, the functional dependence
(\ref{2.14}) can be made local in space through an expansion in
gradients of the hydrodynamic fields. To generate it, $f_{i}$ is
written as a series expansion in a formal parameter $\delta$
measuring the nonuniformity of the system,
\begin{equation}
f_{i}=f_{i}^{(0)}+\delta \,f_{i}^{(1)}+\delta^2
\,f_{i}^{(2)}+\cdots \;, \label{2.15}
\end{equation}
where each factor of $\delta$ means an implicit gradient of a
hydrodynamic field. The local reference states $f_{i}^{(0)}$ are
chosen such that they verify Eqs.\ (\ref{2.4}) and (\ref{2.5}), or
equivalently, the remainder of the expansion must obey the
orthogonality conditions
\begin{equation}
\label{2.16bis}
 \int d{\bf v}\left[ f_{i}({\bf
v})-f_{i}^{(0)}({\bf v})\right] =0\;,\
\end{equation}
\begin{equation}
\sum_{i=1}^2\,m_i\,\int d{\bf v}\,{\bf v}\left[ f_{i}({\bf
v})-f_{i}^{(0)}({\bf v})\right] ={\bf 0}\;,  \label{2.16}
\end{equation}
\begin{equation}
\sum_{i=1}^2\,\frac{m_{i}}{2}\,\int d{\bf v}\,V^{2}\left[
f_{i}({\bf v})-f_{i}^{(0)}({\bf v})\right] =0\;.  \label{2.17}
\end{equation}
The time derivatives of the fields are also expanded as $\partial
_{t}=\partial _{t}^{(0)}+\epsilon \partial _{t}^{(1)}+\cdots $.
The action of the operators $\partial _{t}^{(k)}$ can be obtained
from the balance equations (\ref{2.9})--(\ref{2.11}) when one
takes into account the corresponding expansions for the fluxes and
the cooling rate. This is the usual CE method \cite{CC70} for
solving kinetic equations. The main difference in the case of
inelastic collisions is that the reference state has a time
dependence associated with the cooling that is not proportional to
the gradients. As a consequence, terms from the time derivative
$\partial_t^{(0)}$ are not zero.  In addition, the different
approximations $f_i^{(k)}$ are well-defined functions of the
coefficients of restitution $\alpha_{ij}$, regardless of the
applicability of the corresponding hydrodynamic equations
truncated at that order.

\subsection{Zeroth-order approximation}

To zeroth order in the gradients, Eq.\ (\ref{2.1}) becomes
\begin{equation}
\label{2.17.1}
\partial_t^{(0)}f_i^{(0)}=\sum_{j=1}^2\;
J_{ij}[f_i^{(0)},f_j^{(0)}],
\end{equation}
where use has been made of the fact that gravity is assumed to be
of first order in the uniformity parameter $\delta$. The balance
equations to this order give
\begin{equation}
\label{2.17.2}
\partial_t^{(0)}x_1=\partial_t^{(0)}u_\ell=0,\quad
T^{-1}\partial_t^{(0)}T=p^{-1}\partial_t^{(0)}p=-\zeta^{(0)},
\end{equation}
where $\zeta^{(0)}$ is determined by Eqs.\ (\ref{2.7.1}) and
(\ref{2.7.2}) to zeroth order in the gradients. Since $f_i^{(0)}$
is a normal solution, then the time derivative in Eq.\
(\ref{2.17.1}) can be written as
\begin{equation}
\label{2.17.3}
\partial_t^{(0)}f_i^{(0)}=-\zeta^{(0)}\left(T\partial_T+p\partial_p\right)f_i^{(0)}=
\frac{1}{2}\zeta^{(0)}\frac{\partial}{\partial {\bf V}}\cdot
\left({\bf V}f_i^{(0)}\right).
\end{equation}
The second equality in Eq.\ (\ref{2.17.3}) follows from
dimensional analysis which requires that the dependence of
$f_i^{(0)}$ on $p$ and $T$ is of the form
\begin{equation}
\label{2.17.4} f_i^{(0)}(
V)=x_i\frac{p}{T}v_0^{-d}\Phi_i\left(V/v_0\right),
\end{equation}
where $v_0(t)=\sqrt{2T(m_1+m_2)/m_1m_2}$ is a thermal velocity
defined in terms of the temperature $T(t)$ of the mixture and
$\Phi_i$ is a dimensionless function of the reduced velocity
$V/v_0$. The dependence of $f_i^{(0)}$ on the magnitude of ${\bf
V}$ follows from the isotropy of the zeroth-order distribution
with respect to the peculiar velocity. Thus, the Boltzmann
equation at this order reads
\begin{equation}
\label{2.17.5} \frac{1}{2}\zeta^{(0)}\frac{\partial}{\partial {\bf
V}}\cdot \left({\bf V}f_i^{(0)}\right) \sum_{j=1}^2\;
J_{ij}[f_i^{(0)},f_j^{(0)}].
\end{equation}
Equation (\ref{2.17.5}) has the same form as the Boltzman equation
for a strictly {\em homogeneous} state. The latter is called the
homogeneous cooling state (HCS). \cite{GD99b} Here, however, the
state is not homogeneous because of the requirements
(\ref{2.16bis})--(\ref{2.17}). Instead it is a {\em local} HCS. It
must be emphasized that the presence of this local HCS as the
ground or reference state is not an assumption of the CE expansion
but rather a consequence of the kinetic equations at zeroth order
in the gradient expansion.

The local HCS distribution is the solution of the Boltzmann
equation (\ref{2.17.5}). However, its explicit form is not exactly
known even in the one-component case.\cite{NE98} An accurate
approximation for the zeroth-order solution $f_i^{(0)}$ can be
obtained by using low order truncation of a Sonine polynomial
expansion. The results show that in general, $f_i^{(0)}$ is close
to a Maxwellian at the temperature for that species. Further
details of this solution for hard spheres ($d=3$) can be found in
Ref.\ \onlinecite{GD99b}. An important consequence is that the
kinetic temperatures of each species are different for inelastic
collisions and, consequently the total energy is not equally
distributed between both species (breakdown of energy
equipartition). This violation of energy equipartition has been
confirmed by computer simulation studies \cite{computer} as well
as by real experiments. \cite{exp1} The condition that $f_i^{(0)}$
is {\em normal} in the sense of Eq.\ (\ref{2.14}) (namely, it
depends on time only through its functional dependence on $T$ and
$p$) implies that the ratio $T_i/T\equiv \gamma_i(x_1)$ depends on
the hydrodynamic state through the concentration $x_1$.

The dependence of the temperature ratio $\gamma\equiv
\gamma_1/\gamma_2=T_1/T_2$ on the parameters of the mixture is
obtained by requiring that the partial cooling rates
$\zeta_i^{(0)}$ must be equal,\cite{GD99b} i.e.,
\begin{equation}
\label{2.17bis} \zeta_1^{(0)}=\zeta_2^{(0)}=\zeta^{(0)}.
\end{equation}
These partial cooling rates are nonlinear functionals of the
distributions $f_i^{(0)}$, which are not exactly known. However,
to get the temperature ratio, they can be well estimated by using
Maxwellians at different temperatures:
\begin{equation}
\label{2.18} f_i^{(0)}({\bf V})\to f_{i,M}({\bf
V})=n_i\left(\frac{m_i}{2\pi T_i}\right)^{d/2}\exp\left(-
\frac{m_i V^2}{2T_i}\right).
\end{equation}
In this approximation, one gets \cite{G02}
\begin{eqnarray}
\label{2.19}
\zeta_i^{(0)}&=&\sum_{j=1}^2\zeta_{ij}^{(0)}=\frac{4\pi^{(d-1)/2}}{d\Gamma\left(\frac{d}{2}\right)}
v_0\sum_{j=1}^2
n_j\mu_{ji}\sigma_{ij}^{d-1}\left(\frac{\theta_i+\theta_j}
{\theta_i\theta_j}\right)^{1/2}\nonumber\\
& &\times (1+\alpha_{ij})
\left[1-\frac{\mu_{ji}}{2}(1+\alpha_{ij})
\frac{\theta_i+\theta_j}{\theta_j}\right],
\end{eqnarray}
where
\begin{equation}
\label{2.20} \theta_i=\frac{m_i}{\gamma_i}\sum_{j=1}^2\,m_j^{-1}.
\end{equation}
It must be remarked that the fact that $T_1(t)\neq T_2(t)$ does
not mean that there are additional hydrodynamic degrees of freedom
since the partial temperatures $T_i$ can be expressed in terms of
the granular temperature $T$ as
\begin{equation}
\label{2.20.0} T_1(t)=\frac{\gamma}{1+x_1(\gamma-1)}T(t),\quad
T_2(t)=\frac{1}{1+x_1(\gamma-1)}T(t).
\end{equation}
Note that the reference Maxwellians (\ref{2.18}) for the two
species can be quite different due to the temperature differences.
This contrasts with the ground state considered in more standard
derivations \cite{mixture,SGNT06} where $f_i^{(0)}$ is replaced by
a Maxwellian defined at the same temperature $T$, i.e.,
\begin{equation}
\label{2.20.1} f_i^{(0)}({\bf V})\to n_i\left(\frac{m_i}{2\pi
T}\right)^{d/2}\exp\left(- \frac{m_i V^2}{2T}\right).
\end{equation}
As will show later, the approaches (\ref{2.18}) and (\ref{2.20.1})
lead to different results for the NS transport coefficients.
\begin{figure}[htbp]
\begin{center}
\resizebox{7cm}{!}{\includegraphics{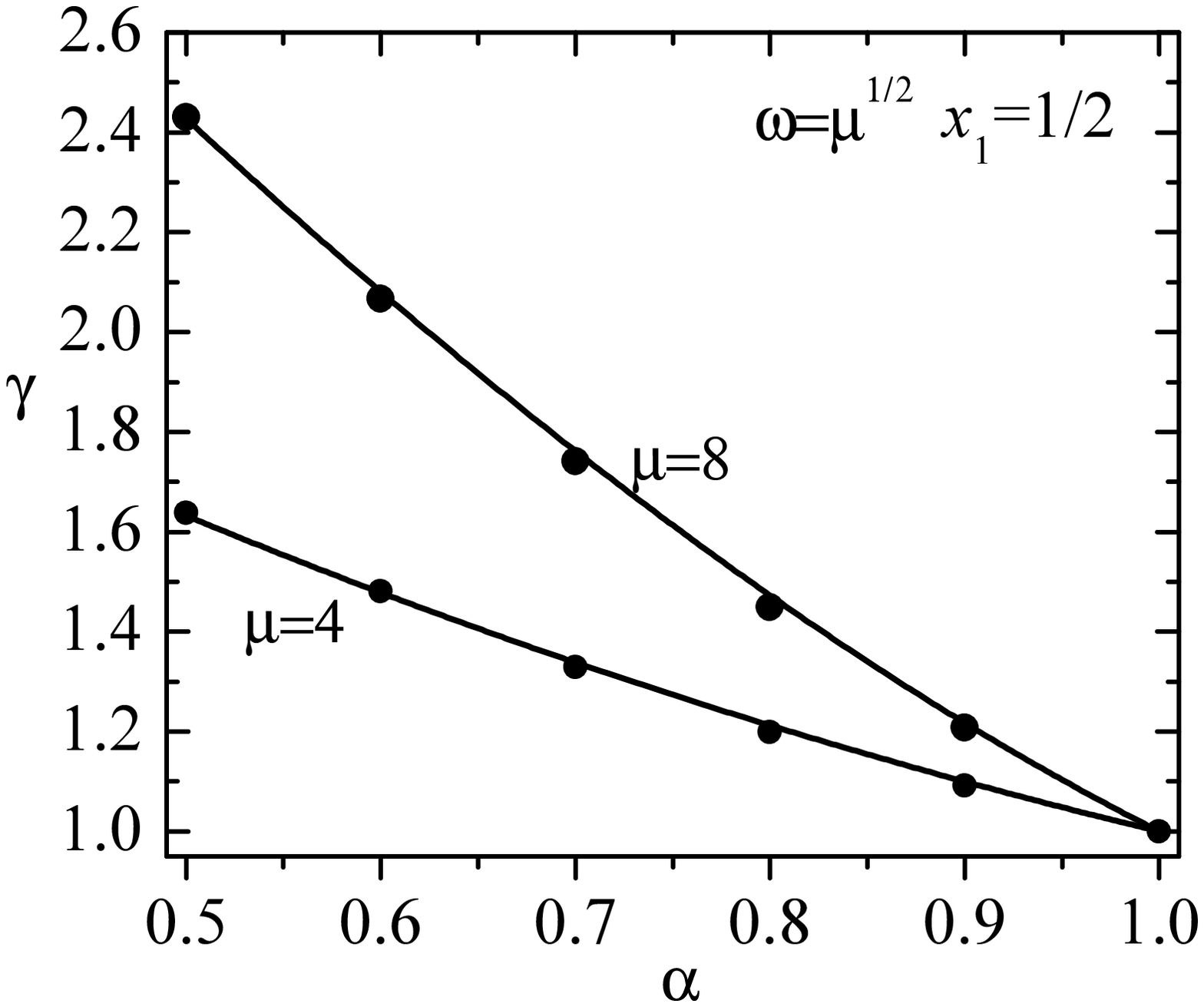}}
\end{center}
\caption{Temperature ratio $\gamma\equiv T_1/T_2$ versus the
(common) coefficient of restitution $\alpha$ for an equimolar
mixture ($x_1=1/2$) of hard disks ($d=2$) with $\omega=\mu^{1/2}$.
Two different values of the mass ratio are considered: $\mu=4$ and
$\mu=8$.  The symbols refer to DSMC results while the lines
represent the theoretical results obtained from the condition
(\ref{2.17bis}).} \label{fig1}
\end{figure}

The solution to Eq.\ (\ref{2.17bis}) gives the temperature ratio
$T_1/T_2$ for any dimension $d$ as a function of the mole fraction
$x_1$, the mass ratio $\mu\equiv m_1/m_2$, the size ratio
$\omega\equiv \sigma_1/\sigma_2$, and the coefficients of
restitution $\alpha_{ij}$. To illustrate the violation of
equipartition theorem, in Fig.\ \ref{fig1} we plot the temperature
ratio versus the coefficient of restitution $\alpha$ for hard
disks ($d=2$) in the case of an equimolar mixture
$x_1=\frac{1}{2}$ for two different mixtures composed by particles
of the same material: $\mu=4$, $\omega=2$, and $\mu=8$,
$\omega=\sqrt{8}$. For the sake of simplicity, we have taken a
common coefficient of restitution $\alpha\equiv
\alpha_{11}=\alpha_{22}=\alpha_{12}$. We also include the
simulation data obtained by solving  numerically the Boltzmann
equation by means of the DSMC method. \cite{Bird} The excellent
agreement between theory and simulation shows the accuracy of the
estimate (\ref{2.19}) to compute the temperature ratio from the
equality of cooling rates (\ref{2.17bis}). We also observe that
the deviations from the energy equipartition increase as the
mechanical differences between the particles of each species
increase.

\section{Navier-Stokes transport coefficients}
\label{sec3}

The CE procedure allows one to determine the NS transport
coefficients of the mixture in the first order of the expansion.
The analysis to first order in gradients is similar to the one
worked out in Ref.\ \onlinecite{GD02} for $d=3$. Here, we only
present the final results with some technical details being given
in Appendix \ref{appA}. The mass, momentum, and heat fluxes are
given, respectively, by
\begin{equation}
{\bf j}_{1}^{(1)}=-\frac{m_{1}m_{2}n}{\rho } D\nabla x_{1}-\frac{
\rho }{p}D_{p}\nabla p-\frac{\rho }{T}D^{\prime }\nabla
T,\hspace{0.3in}{\bf j}_{2}^{(1)}=-{\bf j}_{1}^{(1)},  \label{3.1}
\end{equation}
\begin{equation}
P_{k \ell}^{(1)}=p\;\delta _{k\ell}-\eta \left( \nabla _{\ell
}u_{k}+\nabla _{k}u_{\ell}-\frac{2}{d}\delta _{k\ell} {\bf \nabla
\cdot u}\right),  \label{3.2}
\end{equation}
\begin{equation}
{\bf q}^{(1)}=-T^{2}D^{\prime \prime }\nabla x_{1}-L\nabla
p-\lambda \nabla T. \label{3.3}
\end{equation}
The transport coefficients in these equations are the diffusion
coefficient $D$, the pressure diffusion coefficient $D_p$, the
thermal diffusion coefficient $D'$, the shear viscosity $\eta$,
the Dufour coefficient $D''$, the pressure energy coefficient $L$,
and the thermal conductivity $\lambda$. These coefficients are
defined as
\begin{equation}
D=-\frac{\rho }{dm_{2}n}\int d{\bf v}\,{\bf V}\cdot {\boldsymbol
{\cal A}}_{1}, \label{3.4}
\end{equation}
\begin{equation}
D_{p}=-\frac{m_{1}p}{d\rho}\int d{\bf v}\,{\bf V}\cdot
{\boldsymbol {\cal B}}_{1}, \label{3.5}
\end{equation}
\begin{equation}
D^{\prime }=-\frac{m_{1}T}{d\rho }\int d{\bf v}\,{\bf V}\cdot
{\boldsymbol {\cal C}}_{1}, \label{3.6}
\end{equation}
\begin{equation}
\eta =-\frac{1}{(d-1)(d+2)}\sum_{i=1}^2\,m_i\,\int d{\bf v}\, {\bf
V}{\bf V}:{\boldsymbol {\cal D}}_{i}, \label{3.7}
\end{equation}
\begin{equation}
D^{\prime \prime
}=-\frac{1}{dT^{2}}\sum_{i=1}^2\,\frac{m_i}{2}\,\int d{\bf
v}\,V^{2}{\bf V}\cdot {\boldsymbol {\cal A}}_{i}, \label{3.8}
\end{equation}
\begin{equation}
L=-\frac{1}{d}\sum_{i=1}^2\,\frac{m_i}{2}\,\int d{\bf
v}\,V^{2}{\bf V} \cdot \,{\boldsymbol {\cal B}}_{i}, \label{3.9}
\end{equation}
\begin{equation}
\lambda =-\frac{1}{d}\sum_{i=1}^2\,\frac{m_i}{2}\,\int d{\bf
v}\,V^{2} {\bf V}\cdot {\boldsymbol {\cal C}}_{i}. \label{3.9bis}
\end{equation}
As for ordinary gases, \cite{CC70} the unknowns ${\boldsymbol
{\cal A}}_{i}({\bf V})$, ${\boldsymbol {\cal B}}_{i}({\bf V})$,
${\boldsymbol {\cal C}}_{i}({\bf V})$, and ${\boldsymbol {\cal
D}}_{i}({\bf V})$ are the solutions of the following set of
coupled linear integral equations:
\begin{subequations}
\begin{equation}
\left[ -\zeta ^{(0)}\left( T\partial _{T}+p\partial _{p}\right)
+{\cal L}_{1} \right] {\boldsymbol {\cal A}}_{1}+{\cal
M}_{1}{\boldsymbol {\cal A}}_{2}={\bf A}_{1}+\left( \frac{\partial
\zeta ^{(0)}}{\partial x_{1}}\right) _{p,T}\left( p{\boldsymbol
{\cal
 B}}_{1}+T{\boldsymbol {\cal C}}_{1}\right) ,  \label{3.10a}
\end{equation}
\begin{equation}
\left[ -\zeta ^{(0)}\left( T\partial _{T}+p\partial _{p}\right)
+{\cal L}_{2} \right] {\boldsymbol {\cal A}}_{2}+{\cal
M}_{2}{\boldsymbol {\cal A}}_{1}={\bf A}_{2}+\left( \frac{\partial
\zeta ^{(0)}}{\partial x_{1}}\right) _{p,T}\left( p{\boldsymbol
{\cal
 B}}_{2}+T{\boldsymbol {\cal C}}_{2}\right) ,  \label{3.10b}
\end{equation}
\label{3.10}
\end{subequations}
\begin{subequations}
\begin{equation}
\left[ -\zeta ^{(0)}\left( T\partial _{T}+p\partial _{p}\right)
+{\cal L}_{1}-2\zeta ^{(0)}\right] {\boldsymbol {\cal
B}}_{1}+{\cal M}_{1}{\boldsymbol {\cal B}}_{2}= {\bf
B}_{1}+\frac{T\zeta ^{(0)}}{p}{\boldsymbol {\cal C}}_{1},
\label{3.11a}
\end{equation}
\begin{equation}
\left[ -\zeta ^{(0)}\left( T\partial _{T}+p\partial _{p}\right)
+{\cal L} _{2}-2\zeta ^{(0)}\right] {\boldsymbol {\cal
B}}_{2}+{\cal M}_{2}{\boldsymbol {\cal B}}_{1}= {\bf
B}_{2}+\frac{T\zeta ^{(0)}}{p}{\boldsymbol {\cal C}}_{2},
\label{3.11b}
\end{equation}
\label{3.11}
\end{subequations}
\begin{subequations}
\begin{equation}
\left[ -\zeta ^{(0)}\left( T\partial _{T}+p\partial _{p}\right)
+{\cal L}_{1}-\frac{1}{2}\zeta ^{(0)}\right] {\boldsymbol {\cal
C}}_{1}+{\cal M}_{1} {\boldsymbol {\cal C}}_{2}= {\bf
C}_{1}-\frac{p\zeta ^{(0)}}{2T}{\boldsymbol {\cal B}}_{1},
\label{3.12a}
\end{equation}
\begin{equation}
\left[ -\zeta ^{(0)}\left( T\partial _{T}+p\partial _{p}\right)
+{\cal L}_{2}-\frac{1}{2}\zeta ^{(0)}\right] {\boldsymbol {\cal
C}}_{2}+{\cal M}_{2} {\boldsymbol {\cal C}}_{1}= {\bf
C}_{2}-\frac{p\zeta ^{(0)}}{2T}{\boldsymbol {\cal B}}_{2},
\label{3.12b}
\end{equation}
\label{3.12}
\end{subequations}
\begin{subequations}
\begin{equation}
\label{3.12.0a} \left[ -\zeta ^{(0)}\left( T\partial
_{T}+p\partial _{p}\right) +{\cal L}_{1}\right] {\boldsymbol {\cal
D}}_{1} +{\cal M}_{1}{\boldsymbol {\cal D}}_{2}={\sf D}_1,
\end{equation}
\begin{equation}
\label{3.12.0b} \left[ -\zeta ^{(0)}\left( T\partial
_{T}+p\partial _{p}\right) +{\cal L}_{2}\right] {\boldsymbol {\cal
D}}_{2} +{\cal M}_{2}{\boldsymbol {\cal D}}_{1}={\sf D}_2.
\end{equation}
\label{3.12.0}
\end{subequations}
In the above equations, the quantities ${\bf A}_{i}$, ${\bf
B}_{i}$, ${\bf C}_{i}$, and ${\sf D}_{i}$ are given by Eqs.\
(\ref{a7})--(\ref{a10}), respectively. They depend on the local
HCS distribution $f_i^{(0)}$. In addition, we have introduced the
linearized Boltzmann collision operators
\begin{equation}
{\cal L}_{1}X=-\left( J_{11}[f_{1}^{(0)},X]+J_{11}[X,f_{1}^{(0)}]+
J_{12}[X,f_{2}^{(0)}]\right) \;, \label{3.12.1}
\end{equation}
\begin{equation}
{\cal M}_{1}X=-J_{12}[f_{1}^{(0)},X].  \label{3.12.2}
\end{equation}
The corresponding expressions for the operators ${\cal L}_{2}$ and
${\cal M}_{2}$ can be easily obtained from Eqs.\ (\ref{3.12.1})
and (\ref{3.12.2}) by just making the changes $1\leftrightarrow
2$. Note that in Eq.\ (\ref{3.10}) the cooling rate $\zeta^{(0)}$
depends on $x_1$ explicitly and through its dependence on $\gamma
(x_1)$. This dependence gives rise to significant new
contributions to the integral equations for the transport
coefficients. Furthermore, the external field does not occur in
Eqs.\ (\ref{3.10})--(\ref{3.12.0}). This is because the particular
form of the gravitational force.

\subsection{Sonine polynomial approximation}

So far, all the results are exact. However, explicit expressions
for the NS transport coefficients requires to solve Eqs.\
(\ref{3.10})--(\ref{3.12.0}) as well as the integral equations
(\ref{2.17.5}) for the reference distributions $f_i^{(0)}$. As
said before, the results obtained in the HCS \cite{GD99b} have
shown that $f_i^{(0)}$ is well represented by its Maxwellian form
(\ref{2.18}) in the region of thermal velocities. For this reason
and to provide simple but accurate expressions for the transport
coefficients, non-Gaussian corrections to $f_i^{(0)}$ will be
neglected in our theory. The full expressions for the transport
coefficients in the case $d=3$ (including non-Gaussian
corrections) can be found in Ref.\ \onlinecite{GD02}. It must
remarked that while the effect of these non-Gaussian corrections
on the transport coefficients is not important in the case of the
mass flux and the pressure tensor, the same does not happen for
the heat flux, where the influence of them is not negligible at
high inelasticity. \cite{MGD06} With respect to the functions
$\left( {\boldsymbol {\cal A}}_{i}, {\boldsymbol {\cal B}}_{i},
{\boldsymbol {\cal C}}_{i}, {\boldsymbol {\cal D}}_{i} \right)$,
we will expand them in a series expansion of Sonine polynomials
and will consider only the leading terms. The procedure is
described in Appendix \ref{appB} and only the final expressions
will be provided here.

\subsection{Mass flux}

In dimensionless form, the transport coefficients associated with
the mass flux, $D$, $D_p$, and $D'$ can be written as
\begin{equation}
D=\frac{\rho T}{m_{1}m_{2}\nu _{0}}D^{\ast },\quad
D_{p}=\frac{nT}{\rho \nu _{0}}D_{p}^{\ast },\quad D^{\prime
}=\frac{nT}{\rho \nu _{0}}D^{\prime}{}^{\ast },  \label{3.n1}
\end{equation}
where $\nu _{0}=n\sigma _{12}^{d-1}v_{0}$ is an effective
collision frequency. The explicit forms are
\begin{equation}
D^{\ast}=\left( \nu ^{\ast }-\frac{1}{2}\zeta ^{\ast
}\right)^{-1}\left[ \left( \frac{\partial }{\partial
x_{1}}x_{1}\gamma _{1}\right) _{p,T}+\left( \frac{\partial \zeta
^{\ast }}{\partial x_{1}} \right) _{p,T}\left( 1-\frac{\zeta
^{\ast}}{2\nu ^{\ast }}\right) D_{p}^{\ast }\right] , \label{3.13}
\end{equation}
\begin{equation}
D_{p}^{\ast }=x_{1}\left( \gamma _{1}-\frac{\mu}{x_2+\mu x_1}
\right) \left( \nu ^{\ast }-\frac{3}{2}\zeta ^{\ast }+\frac{\zeta
^{\ast 2}}{ 2\nu ^{\ast }}\right) ^{-1}, \label{3.14}
\end{equation}
\begin{equation}
D^{\prime\ast }=-\frac{\zeta ^{\ast }}{2\nu ^{\ast }}D_{p}^{\ast
}. \label{3.15}
\end{equation}
Here, $\zeta^*=\zeta^{(0)}/\nu_0$, and $\nu^*$ is given by
\begin{equation}
\label{3.16}
\nu^*=\frac{2\pi^{(d-1)/2}}{d\Gamma\left(\frac{d}{2}\right)}
(1+\alpha_{12})
\left(\frac{\theta_1+\theta_2}{\theta_1\theta_2}\right)^{1/2}\left(x_2\mu_{21}+
x_1\mu_{12}\right).
\end{equation}
Since ${\bf j}_{1}^{(1)}=-{\bf j}_{2}^{(1)}$ and $\nabla
x_{1}=-\nabla x_{2}$, $D^*$ must be symmetric while $D_{p}^*$ and
$D^{\prime \ast}$ must be antisymmetric with respect to the
exchange $1\leftrightarrow 2$ . This can be easily verified by
noting that $x_{1}\gamma_{1}+x_{2}\gamma_{2}=1$. The expressions
for $D^*$, $D_{p}^*$ and $D^{\prime \ast}$ reduce to those
recently obtained \cite{BRM05} in the tracer limit ($x_1\to 0$).

\subsection{Pressure tensor}

The shear viscosity coefficient $\eta$ can be written as
\begin{equation}
\label{3.17}
\eta=\frac{p}{\nu_0}\left(x_1\gamma_1^2\eta_1^*+x_2\gamma_2^2\eta_2^*\right),
\end{equation}
where the expression of the (dimensionless) partial contribution
$\eta_i^*$ is
\begin{equation}
\label{3.18}
\eta_1^*=2\frac{\gamma_2(2\tau_{22}-\zeta^{*})-2\gamma_1\tau_{12}}
{\gamma_1\gamma_2[\zeta^*-2\zeta^{*}
(\tau_{11}+\tau_{22})+4(\tau_{11}\tau_{22}-\tau_{12}\tau_{21})]}.
\end{equation}
Here, we have introduced the (reduced) collision frequencies
$\tau_{11}$ and $\tau_{12}$ given by
\begin{widetext}
\begin{eqnarray}
\label{3.19}
\tau_{11}&=&\frac{2\pi^{(d-1)/2}}{d(d+2)\Gamma\left(\frac{d}{2}\right)}\left\{
x_1\left(\frac{\sigma_{1}}{\sigma_{12}}\right)^{d-1}(2\theta_1)^{-1/2}(3+2d-3\alpha_{11})
(1+\alpha_{11})\right.\nonumber\\
& &+2x_2 \mu_{21}(1+\alpha_{12}) \theta_1^{3/2}\theta_2^{-1/2}
\left[
(d+3)(\mu_{12}\theta_2-\mu_{21}\theta_1)\theta_1^{-2}(\theta_1+\theta_2)^{-1/2}\right.\nonumber\\
& &
\left.\left.+\frac{3+2d-3\alpha_{12}}{2}\mu_{21}\theta_1^{-2}(\theta_1+\theta_2)^{1/2}
+\frac{2d(d+1)-4}{2(d-1)}\theta_1^{-1}(\theta_1+\theta_2)^{-1/2}\right]\right\},
\end{eqnarray}
\begin{eqnarray}
\label{3.20}
\tau_{12}&=&\frac{4\pi^{(d-1)/2}}{d(d+2)\Gamma\left(\frac{d}{2}\right)}
x_2\frac{\mu_{21}^2}{\mu_{12}}\theta_1^{3/2}\theta_2^{-1/2}
(1+\alpha_{12})\nonumber\\
& \times&\left[
(d+3)(\mu_{12}\theta_2-\mu_{21}\theta_1)\theta_2^{-2}(\theta_1+\theta_2)^{-1/2}\right.\nonumber\\
& &
\left.+\frac{3+2d-3\alpha_{12}}{2}\mu_{21}\theta_2^{-2}(\theta_1+\theta_2)^{1/2}
-\frac{2d(d+1)-4}{2(d-1)}\theta_2^{-1}(\theta_1+\theta_2)^{-1/2}\right].
\end{eqnarray}
\end{widetext}
A similar expression can be obtained for $\eta_2^*$ by just making
the changes $1 \leftrightarrow 2$.

\subsection{Heat flux}

The case of the heat flux is more involved since it requires to
consider the second Sonine approximation. The transport
coefficients appearing in the heat flux, $D''$, $L$, and $\lambda$
can be written as
\begin{equation}
D^{\prime \prime
}=-\frac{d+2}{2}\frac{n}{(m_{1}+m_{2})\nu_0}\left[ \frac{
x_{1}\gamma _{1}^{3}}{\mu _{12}}d_{1}^*+\frac{x_{2}\gamma
_{2}^{3}}{\mu _{21}}d_{2}^*-\left( \frac{\gamma _{1}}{\mu
_{12}}-\frac{\gamma _{2}}{\mu _{21}}\right) D^{\ast }\right] ,
\label{4.7}
\end{equation}
\begin{equation}
L=-\frac{d+2}{2}\frac{T}{(m_{1}+m_{2})\nu_0}\left[
\frac{x_{1}\gamma _{1}^{3}}{\mu _{12}}\ell
_{1}^*+\frac{x_{2}\gamma _{2}^{3}}{\mu _{21}}\ell _{2}^*-\left(
\frac{\gamma _{1}}{\mu _{12}}-\frac{\gamma _{2}}{\mu _{21}}
\right) D_{p}^{\ast }\right] ,  \label{4.8}
\end{equation}
\begin{equation}
\lambda =-\frac{d+2}{2}\frac{nT}{(m_{1}+m_{2})\nu _{0}}\left[
\frac{ x_{1}\gamma _{1}^{3}}{\mu _{12}}\lambda
_{1}^*+\frac{x_{2}\gamma _{2}^{3}}{\mu_{21}}\lambda _{2}^*-\left(
\frac{\gamma _{1}}{\mu _{12}}-\frac{\gamma _{2}}{ \mu
_{21}}\right) D^{\prime }{}^{\ast }\right] , \label{4.9}
\end{equation}
where the coefficients $D^{\ast}$, $D_{p}^{\ast }$, and
 $D^{\prime \ast }$ are given by
 Eqs.\ (\ref{3.13})--(\ref{3.15}), respectively. The expressions
 of the (dimensionless) Sonine coefficients $d_{i}^{*}$, $\ell
_{i}^*$, and $\lambda _{i}^*$ are
\begin{eqnarray}
\label{4.n2} d_1^{*}&=&\frac{1}{\Delta}\left\{2\left[2
\nu_{12}Y_2-Y_1(2\nu_{22}-3\zeta^*)\right]\left[\nu_{12}\nu_{21}-\nu_{11}\nu_{22}
+2(\nu_{11}+\nu_{22})\zeta^*-4\zeta^{*2}\right]\right.\nonumber\\
& &+2\left( \frac{\partial \zeta ^{\ast }}{\partial x_{1}}\right)
_{p,T}(Y_3+Y_5)\left[2\nu_{12}\nu_{21}+2\nu_{22}^2-\zeta^*(
7\nu_{22}-6\zeta^{*})\right]\nonumber\\
& & \left. -2\nu_{12}\left( \frac{\partial \zeta ^{\ast
}}{\partial
x_{1}}\right)_{p,T}(Y_4+Y_6)\left(2\nu_{11}+2\nu_{22}-7\zeta^*\right)\right\},
\end{eqnarray}
\begin{eqnarray}
\label{4.n3} \ell_1^*&=&\frac{1}{\Delta}\left\{-2Y_3\left[2
(\nu_{12}\nu_{21}-\nu_{11}\nu_{22})\nu_{22}+\zeta^*(7\nu_{11}\nu_{22}-5\nu_{12}\nu_{21}+2\nu_{22}^2
-6\nu_{11}\zeta^*-7\nu_{22}\zeta^*+6\zeta^{*2})\right]\right.\nonumber\\
&&
+2Y_4\nu_{12}\left[2\nu_{12}\nu_{21}-2\nu_{11}\nu_{22}+2\zeta^*(\nu_{11}+\nu_{22})
-\zeta^{*2}\right]\nonumber\\
& +&
\left.2Y_5\zeta^*\left[2\nu_{12}\nu_{21}+\nu_{22}(2\nu_{22}-7\zeta^*)+6\zeta^{*2}\right]
-2\nu_{12}\zeta^*Y_6\left[2(\nu_{11}+\nu_{22})-7\zeta^*\right]
\right\},
\end{eqnarray}
\begin{eqnarray}
\label{4.n4}
\lambda_1^*&=&\frac{1}{\Delta}\left\{-Y_3\zeta^*\left[2
\nu_{12}\nu_{21}+\nu_{22}(2\nu_{22}-7\zeta^*)+6\zeta^{*2}\right]+\nu_{12}\zeta^*
Y_4\left[2(\nu_{11}+\nu_{22})-7\zeta^*\right] \right. \nonumber\\
&
&-Y_5\left[4\nu_{12}\nu_{21}(\nu_{22}-\zeta^*)+2\nu_{22}^2(5\zeta^*-2\nu_{11})+2\nu_{11}
(7\nu_{22}\zeta^*-6\zeta^{*2})+5\zeta^{*2}(6\zeta^*-7\nu_{22})\right]\nonumber\\
& & \left.
+\nu_{12}Y_6\left[4\nu_{12}\nu_{21}+2\nu_{11}(5\zeta^*-2\nu_{22})+\zeta^*(10\nu_{22}-
23\zeta^*)\right]\right\},
\end{eqnarray}
where
\begin{equation}
\label{4.n5}
\Delta=\left[4(\nu_{12}\nu_{21}-\nu_{11}\nu_{22})+6\zeta^*(\nu_{11}+\nu_{22})-9\zeta^{*2}\right]
\left[\nu_{12}\nu_{21}-\nu_{11}\nu_{22}+2\zeta^*(\nu_{11}+\nu_{22})-4\zeta^{*2}\right].
\end{equation}
In the above equations, the Y's are defined by Eqs.\
(\ref{4.14})--(\ref{4.16}), while the (reduced) collision
frequencies $\nu_{11}$ and $\nu_{12}$ are given by Eqs.\
(\ref{c18}) and (\ref{c19}), respectively. The expressions for
$d_2^{*}$, $\ell_2^*$, and $\lambda_2^*$ can be obtained from
Eqs.\ (\ref{4.n2})--(\ref{4.n5}) by setting $1\leftrightarrow 2$.
As expected, our results for the heat flux show that $D^{\prime
\prime }$ is antisymmetric with respect to the change
$1\leftrightarrow 2$, while $L$ and $ \lambda $ are symmetric.
Consequently, in the case of mechanically equivalent particles
($m_{1}=m_{2}\equiv m$, $\sigma _{1}=\sigma _{2}\equiv \sigma$,
$\alpha _{ij}\equiv \alpha $), the coefficient $D^{\prime \prime}$
vanishes.

In the three-dimensional case ($d=3$), all the above expressions
for the transport coefficients reduce to those previously derived
for hard spheres \cite{GD02,MGD06} when one takes Maxwellian
distributions (\ref{2.18}) for the zeroth-order approximations
$f_i^{(0)}$. For mechanically equivalent particles, the results
obtained by Brey and Cubero \cite{BC01} for a $d$-dimensional
monocomponent gas are also recovered. This confirms the
self-consistency of the results derived here.

\begin{figure}[htbp]
\begin{center}
\resizebox{8cm}{!}{\includegraphics{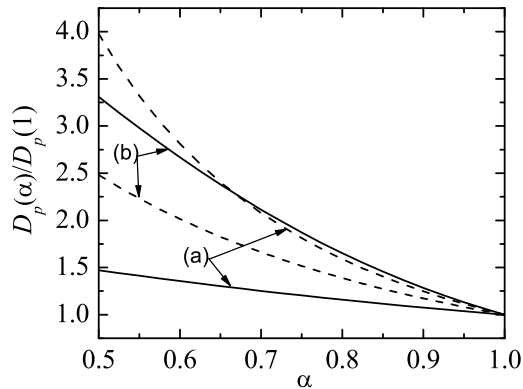}}
\end{center}
\caption{Plot of the reduced pressure diffusion coefficient
$D_p(\alpha)/D_p(1)$ as a function of the (common) coefficient of
restitution $\alpha$ for binary mixtures with $x_1=0.2$,
$\omega=1$ in the case of a three-dimensional system ($d=3$) and
two values of the mass ratio $\mu$: $\mu=0.5$ (a) and  $\mu=4$
(b). The solid lines refer to the results derived here and the
dashed lines correspond to the results assuming the equality of
the partial temperatures.} \label{fig4}
\end{figure}
\begin{figure}[htbp]
\begin{center}
\resizebox{8cm}{!}{\includegraphics{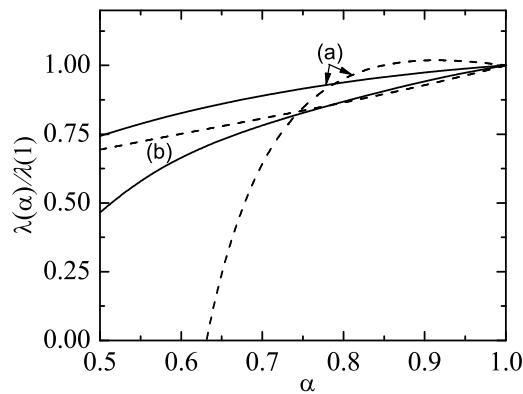}}
\end{center}
\caption{Plot of the reduced thermal conductivity coefficient
$\lambda(\alpha)/\lambda(1)$ as a function of the (common)
coefficient of restitution $\alpha$ for binary mixtures with
$x_1=0.2$, $\omega=1$ in the case of a three-dimensional system
($d=3$) and two values of the mass ratio $\mu$: $\mu=0.5$ (a) and
$\mu=4$ (b). The solid lines refer to the results derived here and
the dashed lines correspond to the results assuming the equality
of the partial temperatures.} \label{fig5}
\end{figure}

\subsection{Comparison with other theories}

Before checking the accuracy of our expressions by comparing them
with computer simulations, it is instructive first to make some
comparison with previous results. \cite{mixture,SGNT06} These
results assume energy equipartition ($T_i=T$) and so, they are
based on a standard CE expansion around the Maxwellian
(\ref{2.20.1}) instead of the local HCS distribution. Figures
\ref{fig4} and \ref{fig5} show the dependence of the reduced
pressure diffusion coefficient $D_p(\alpha)/D_p(1)$ and the
reduced thermal conductivity coefficient
$\lambda(\alpha)/\lambda(1)$, respectively, as a function of the
(common) coefficient of restitution $\alpha_{ij}\equiv \alpha$ for
$d=3$, $\omega=1$, $x_1=0.2$, and two different mass ratios $\mu$:
$\mu=0.5$ (a) and  $\mu=4$ (b). Here, $D_p(1)$ and $\lambda(1)$
are the values of $D_p$ and $\lambda$ for elastic collisions. We
see that the deviation from the functional form for elastic
collisions is quite important in both theories, even for moderate
dissipation. It is apparent that the dependence of the transport
coefficients on dissipation is quantitatively different in both
models, especially at strong dissipation (say for instance,
$\alpha=0.5$). This clearly shows the real quantitative effect of
two different species temperatures on transport in granular
mixtures.

\section{Comparison with Monte Carlo simulations}
\label{sec4}

A said before, the expressions derived in Sec.\ \ref{sec3} for the
NS transport coefficients have been obtained by considering two
different approximations. First, since the deviation of
$f_i^{(0)}$ from its Maxwellian form (\ref{2.18}) is quite small
in the region of thermal velocities, we have used the Maxwellian
distribution (\ref{2.18}) as a trial function for $f_i^{(0)}$.
Second, we have only considered the leading terms of an expansion
of the distribution $f_i^{(1)}$ in Sonine polynomials. Both
approximations allow one to offer a simplified kinetic theory for
a $d$-dimensional granular binary mixture. To check the accuracy
of the above predictions, in this Section we numerically solve the
Boltzmann equation by means of the DSMC method \cite{Bird} and
compare theory and simulation in the cases of the diffusion
coefficient $D$ (in the tracer limit) and the shear viscosity
coefficient $\eta$. Previous comparisons carried out for hard
spheres \cite{GM04,MG03} when one takes into account the
deviations of $f_i^{(0)}$ from their Maxwellian forms have shown
good agreement between theory and simulation, even for strong
dissipation (say $\alpha \gtrsim  0.5$). Here, we expect that such
good agreement is also maintained in the case of hard disks when
one replaces $f_i^{(0)}\to f_{i,M}$. Let us study each coefficient
separately.

\subsection{Tracer diffusion coefficient}

We consider the special case in which one of the components of the
mixture (say, for instance, species $1$) is present in tracer
concentration ($x_1\to 0$). In this situation, $\partial
\zeta^*/\partial x_1 \to 0$ and so, the expression (\ref{3.13})
for the reduced diffusion coefficient $D^*$ becomes
\begin{equation}
\label{4.1} D^*=\frac{\gamma}{\nu^*-\frac{1}{2}\zeta^{*}},
\end{equation}
where now
\begin{equation}
\label{4.2}
\zeta^{*}=\frac{\pi^{(d-1)/2}}{d\Gamma\left(d/2\right)}
\left(\frac{\sigma_2}{\sigma_{12}}\right)^{d-1}\sqrt{2\mu_{12}}
(1-\alpha_{22}^2),
\end{equation}
\begin{equation}
\label{4.3} \nu^*=\frac{2\pi^{(d-1)/2}}{d\Gamma\left(d/2\right)}
\mu_{21} \sqrt{\mu_{12}+\mu_{21}\gamma}
 (1+\alpha_{12}).
\end{equation}

The diffusion coefficient of impurities in a granular gas
undergoing homogeneous cooling state can be measured in simulation
from the mean square displacement of
the tracer particle after a time interval $t$: \cite{M89,GM04}
\begin{equation}
\label{4.4} \frac{\partial}{\partial t}\langle |{\bf r}(t)-{\bf
r}(0)|^2 \rangle =\frac{2dD}{n_2}.
\end{equation}
Equation (\ref{4.4}) is the Einstein form. This relation
(written in appropriate dimensionless variables to eliminate the
time dependence of $D(t)$) was used in Ref.\ \onlinecite {GM04} to
measure the diffusion coefficient for hard spheres. More details
on this procedure can be found in Ref.\ \onlinecite {GM04}.

\begin{figure}[htbp]
\begin{center}
\resizebox{7cm}{!}{\includegraphics{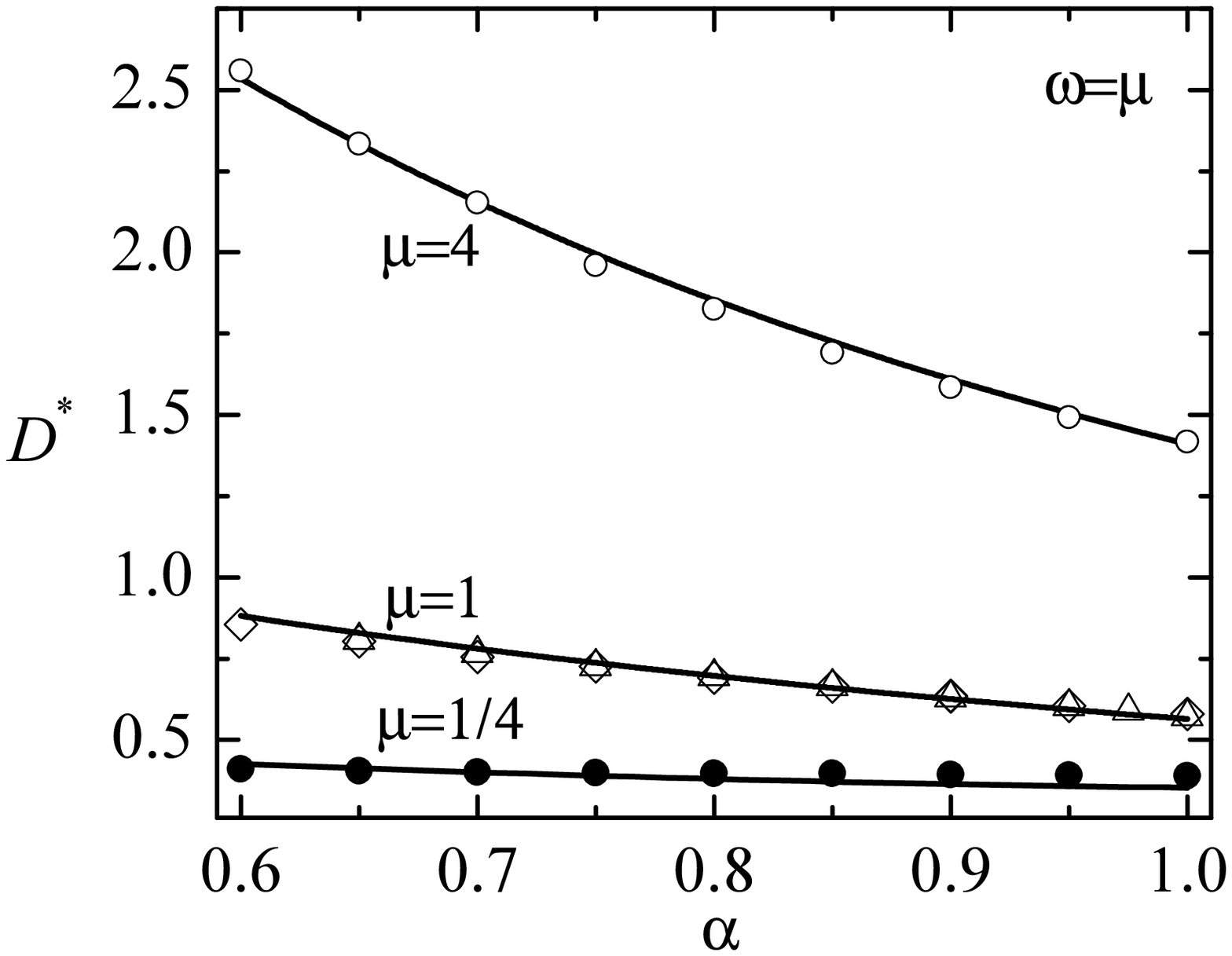}}
\end{center}
\caption{Plot of the reduced diffusion coefficient $D^*$ as a
function of the (common) coefficient of restitution $\alpha$ for
binary mixtures with $\omega=\mu$ in the case of a two-dimensional
system ($d=2$). The symbols are computer simulation results
obtained from the mean square displacement and the lines are the
theoretical results obtained in the first Sonine approximation.
The DSMC results correspond to $\mu=1/4$ ({\Large $\bullet$}),
$\mu=4$ ({\Large $\circ$}) and $\mu=1$ ({\Large $\diamond$}).
Molecular dynamics results reported in Ref.\ \onlinecite{BRCG00}
for $\mu=1$ ($\triangle$) have also been included.} \label{fig2}
\end{figure}

If the hydrodynamic description (or normal solution in the context
of the CE method) applies, then the diffusion coefficient $D(t)$
depends on time only through its dependence on the temperature
$T(t)$. In this case, after a transient regime, the reduced
diffusion coefficient $D^*=(m_1m_2/\rho)D(t)\nu_0(t)/T(t)$
achieves a time-independent value. Here, we compare the steady
state values of $D^*$ obtained from Monte Carlo simulations with
the theoretical predictions given by the first Sonine
approximation (\ref{4.1}). The dependence of $D^*$ on the common
coefficient of restitution $\alpha_{ij}\equiv \alpha$ is shown in
Fig.\ \ref{fig2} in the case of hard disks for three different
systems. The symbols refer to computer simulations while the lines
correspond to the kinetic theory results given by Eq.\
(\ref{4.1}). Molecular dynamics (MD) results reported in Ref.\
\onlinecite{BRCG00} when impurities and particles of the gas are
mechanically equivalent have also been included. We observe that
MD and DSMC results for $\mu=\omega=1$ are consistent among
themselves in the range of values of $\alpha$ explored. This good
agreement gives support to the applicability of the inelastic
Boltzmann equation beyond the quasielastic limit. It is apparent
that the agreement between the first Sonine approximation and
simulation results is excellent when impurities and particles of
the gas are mechanically equivalent and when impurities are much
heavier and/or much larger than the particles of the gas (Brownian
limit). However, some discrepancies between simulation an theory
are found with decreasing values of the mass ratio $m_1/m_2$ and
the size ratio $\sigma_1/\sigma_2$. These discrepancies are not
easily observed in Fig.\ \ref{fig2} because of the small magnitude
of $D^*$ for $\mu=1/4$. The above findings agree with those
previously reported for hard spheres, \cite{GM04} where it was
shown that the second Sonine approximation improves the
qualitative predictions from the first Sonine approximation for
the cases in which the gas particles are heavier and/or larger
than impurities. The comparison carried out here for disks
confirms the above expectations and shows that the Sonine
polynomial expansion exhibits a slow convergence for sufficiently
small values of the mass ratio $\mu$ and/or the size ratio
$\omega$.

\subsection{Shear viscosity coefficient}

The shear viscosity $\eta$ is perhaps the most widely studied
transport coefficient in granular fluids. In the case of granular
mixtures, this coefficient has been measured \cite{MG03} when the
system is {\em heated} by the action of an external driving force
(thermostat) that exactly compensates for cooling effects
associated with dissipation of collisions. The corresponding shear
viscosity of the mixture (which slightly differs from the one
obtained in the free cooling case) has been determined by means of
the CE method in the low-density regime \cite{MG03} as well as for
a moderate dense mixture. \cite{GM03} The theoretical predictions
compare reasonably well with the corresponding numerical solutions
of the Boltzmann and Enskog kinetic equations.

More recently, a new alternative method has been proposed to
measure the (true) NS shear viscosity coefficient. \cite{MSG05}
The method is based on the simple shear flow state modified by the
introduction of a deterministic non-conservative force (which
compensates for the collisional cooling) along with a stochastic
process. While the external force is introduced to allow the
granular fluid to approach a Newtonian regime, the stochastic
process is introduced to mimic the conditions appearing in the CE
method to NS order. Although the method was originally devised to
a single granular gas, its extension to multicomponent systems is
straightforward. Here, we use this procedure to measure the shear
viscosity of the mixture by means of the DSMC method. More
technical details on this procedure and its application to dense
gases can be found in Ref.\ \onlinecite{MSG05}.
\begin{figure}[htbp]
\begin{center}
\resizebox{7cm}{!}{\includegraphics{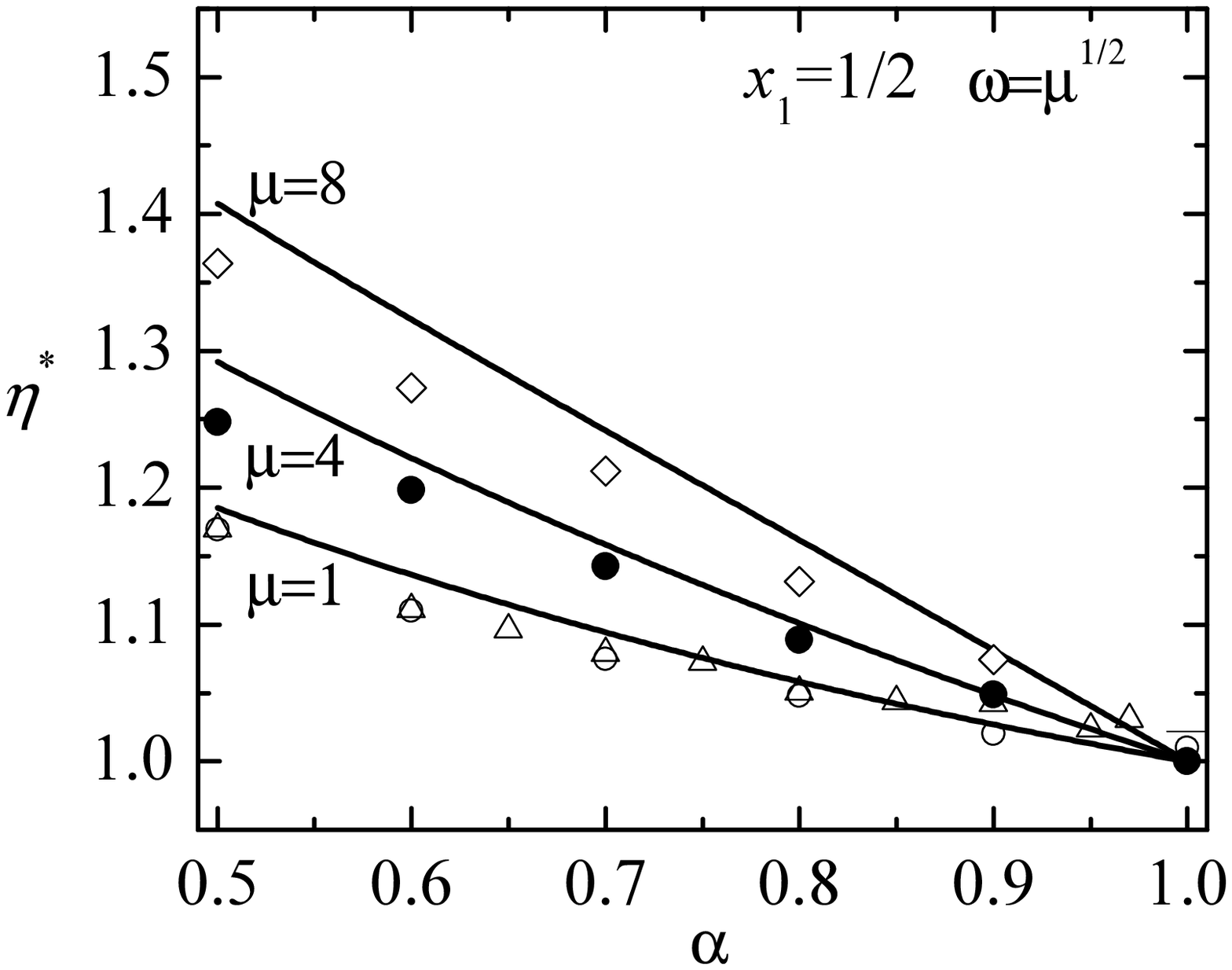}}
\end{center}
\caption{Plot of the reduced shear viscosity
$\eta^*(\alpha)=\eta(\alpha)/\eta(1)$ as a function of the
(common) coefficient of restitution $\alpha$ for binary mixtures
constituted by particles of the same mass density
($\omega=\mu^{1/2}$) in the case of a two-dimensional system
($d=2$). The symbols are computer simulation results and the lines
are the theoretical results obtained in the first Sonine
approximation. The DSMC results correspond to $\mu=1$ ({\Large
$\circ$}), $\mu=4$ ({\Large $\bullet$}) and $\mu=8$ ({\Large
$\diamond$}). We have also included DSMC results obtained in Ref.\
\onlinecite{BRM04} for $\mu=1$ ($\triangle$) from the Green-Kubo
relation.} \label{fig3}
\end{figure}

Comparison between the first Sonine approximation and computer
simulations for $\eta^*(\alpha)=\eta(\alpha)/\eta(1)$ is shown in
Fig.\ \ref{fig3} for three different mixtures constituted by
particles of the same mass density (i.e., $\mu=\omega^d$) in the
case of a two-dimensional system. Here, $\eta(1)$ refers to the
elastic value for the shear viscosity coefficient and we have
assumed again a common value of the coefficient of restitution
$\alpha$. The symbols represent the simulation data and the lines
correspond to the theoretical results. We have also included
recent simulation results \cite{BRM04} for $\eta^*$ obtained from
the Green-Kubo relation in the one-component case. Good agreement
among the data presented here and those reported in Ref.\
\onlinecite{BRM04} for $\mu=1$ is observed. In addition, as
happens for hard spheres, \cite{MG03} we see that in general the
agreement between the first Sonine approximation and simulation is
quite good. At a quantitative level, the theory slightly
overestimates the simulation data, especially for strong
dissipation and for mixtures of particles of different masses
and/or sizes. However, such discrepancies are quite small since
for instance, they are smaller than 3\% at $\alpha=0.5$ for
$\mu=8$ and $\omega=\sqrt{8}$. This shows again the reliability of
the first Sonine approximation for the shear viscosity
coefficient. It must be noted that this conclusion cannot in
principle be extended to the transport coefficients associated with
the heat flux since recent comparisons for a single gas
\cite{BRM04,BRMMG05,MSG06} have shown significant discrepancies between
the first Sonine approximation and computer simulations for high
inelasticity (say $\alpha \lesssim 0.7$). In this case, the
agreement between theory and simulation can be significantly
improved by the use of a modified first Sonine approximation.
\cite{GSM06}

\section{Discussion}
\label{sec5}

The main objective of this work has been to obtain the NS
transport coefficients of a granular binary mixture at low
density. In contrast to previous works, \cite{mixture,SGNT06} the
present study is based on a modified CE solution of the inelastic
Boltzmann equation that takes into account non-equipartition of
energy. There is no phenomenology involved as the equations and
the transport coefficients have been derived systematically from
the inelastic Boltzmann equation by the CE expansion around the
local HCS. Since the spatial gradients are assumed to be
independent of the coefficients of restitution, although the NS
equations restrict their applicability to first order in gradients
the corresponding transport coefficients hold {\em a priori} to
arbitrary degree of inelasticity. All the calculations have been
performed in an arbitrary number $d$ of dimensions, previous
results \cite{GD02} being recovered for $d=3$.

The constitutive equations to NS order for the mass flux, the
stress tensor, and the heat flux are given  by Eqs.\
(\ref{3.1})--(\ref{3.3}), respectively. The associated transport
coefficients are the mutual diffusion coefficient $D$, the
pressure diffusion coefficient $D_p$, and the thermal diffusion
coefficient $D'$ in the case of the mass flux, the shear viscosity
coefficient $\eta$ for the pressure tensor, and the Dufour
coefficient $D''$, the pressure energy coefficient $L$, and the
thermal conductivity $\lambda$ in the case of the heat flux. These
coefficients are determined from the solutions of the set of
coupled linear integral equations (\ref{3.10})--(\ref{3.12.0}). In
addition, the NS transport coefficients also depend on the
reference distributions $f_i^{(0)}$, which are not Maxwellians
because they obey the integral equations (\ref{2.17.5}). To solve
the above integral equations and provide good estimates for the
transport coefficients, we have considered two approximations: (i)
the distributions $f_i^{(0)}$ have been replaced by their
Maxwellian forms (\ref{2.18}) at the temperature $T_i$ for that
species and, (ii) we have only considered the leading terms in a
series of Sonine polynomials for the first-order distribution
$f_i^{(1)}$. By using both approximations, explicit expressions of
the {\em seven} NS transport coefficients have been obtained as
functions of the coefficients of restitution and the concentration
and the ratios of mass and diameters. In dimensionless forms, the
coefficients $D$, $D_p$, and $D'$ are given by Eqs.\
(\ref{3.13})--(\ref{3.15}), respectively, the shear viscosity
$\eta$ is given by Eqs.\ (\ref{3.17}) and (\ref{3.18}), while the
expressions of the coefficients $D''$, $L$, and $\lambda$ are
provided by Eqs.\ (\ref{4.7})--(\ref{4.n5}).

\begin{figure}[htbp]
\begin{center}
\resizebox{7cm}{!}{\includegraphics{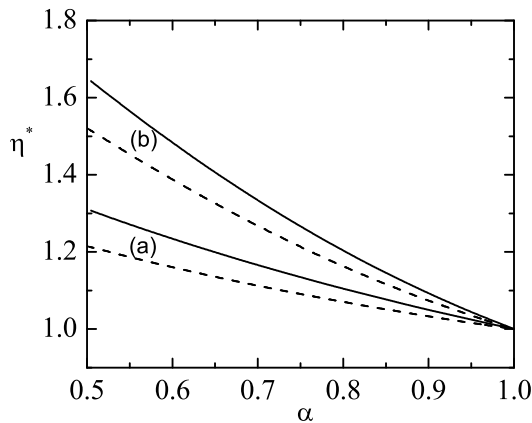}}
\end{center}
\caption{Plot of the reduced shear viscosity coefficient
$\eta^*=\eta(\alpha)/\eta(1)$ as a function of the (common)
coefficient of restitution $\alpha$ for binary mixtures with
$x_1=0.2$, $\omega=1$ and two values of the mass ratio $\mu$:
$\mu=0.5$ (a) and $\mu=4$ (b). The solid lines refer to spheres
($d=3$) while the dashed lines correspond to disks ($d=2$).}
\label{fig6}
\end{figure}

Previous results \cite{mixture} derived from the CE method have
typically introduced additional assumptions for convenience that
are not internally consistent with constructing a solution to the
Boltzmann equation. Thus, in most of the cases the reference state
$f_i^{(0)}$ has been chosen to be a Maxwellian at the same
temperature [see Eq.\ (\ref{2.20.1})]. This assumption is presumed
to give accurate results at weak dissipation where energy
equipartition can be still considered as a good approximation.
However, as shown in Fig.\ \ref{fig1}, the temperature ratio
$T_1/T_2$ clearly differs from 1 as dissipation increases. Here,
we have replaced $f_i^{(0)}\to f_{i,M}$ so that, the influence of
the fourth-cumulants $c_i$ of $f_i^{(0)}$ has been
ignored.\cite{GD99b} Comparison between the expressions derived in
this paper by taking the approximation (\ref{2.18}) with those
obtained by assuming energy equipartition shows important
discrepancies as the coefficient of restitution decreases [see
Figs.\ \ref{fig4} and \ref{fig5}]. Moreover, as an added value of
our theory, the use of the Maxwellian approximation (\ref{2.18})
for $f_i^{(0)}$ allows one to provide simple and explicit
expressions for all the transport coefficients in terms of the
parameters of the mixture. This contrasts with the relatively
recent study for hard spheres \cite{GD02} where the constitutive
relations for the fluxes were not explicitly displayed.

As a complementary route and to check the reliability of our
theory, the analytical results derived for the diffusion
coefficient $D$ and the shear viscosity $\eta$ in the first Sonine
approximation have been compared with those obtained from
numerical solutions of the Boltzmann equation by means of the DSMC
method for a two-dimensional system. For the sake of simplicity,
all the simulations have considered a common coefficient of
restitution $\alpha\equiv \alpha_{ij}$. As expected, theory and
simulation clearly show that the influence of dissipation on mass
and momentum transport is quite important since there is a
relevant dependence of the diffusion $D$ and viscosity $\eta$
coefficients on $\alpha$. With respect to the accuracy of the
theoretical predictions, we see that in general the CE results in
the first Sonine approximation exhibit a good agreement with the
simulation data. Exceptions to this agreement are extreme mass or
size ratios and strong dissipation. These discrepancies are
basically due to the use of the first Sonine approximation and can
be partially mitigated by considering the second and third Sonine
approximations \cite{GM04} or the use of a modified first Sonine
approximation. \cite{GSM06}

As said in the Introduction, the results obtained in this paper
are of great practical interest since most of the experiments and
simulations are performed in two dimensions. On the other hand,
apart from this practical interest, the knowledge of the NS
transport coefficients of a $d$-dimensional mixture allows one to
investigate the influence of dimensionality on the transport
properties of the system. To illustrate this effect, in Fig.\
\ref{fig6} we plot the reduced shear viscosity
$\eta^*\equiv\eta(\alpha)/\eta(1)$ versus the coefficient of
restitution $\alpha$ for $\omega=1$, $x_1=0.2$, and two different
mass ratios $\mu$: $\mu=0.5$ (a) and $\mu=4$ (b). We have
considered the physical cases of hard spheres (solid lines) and
hard disks (dashed lines). Although the qualitative dependence of
$\eta^*$ on $\alpha$ is quite similar in both systems, we observe
that the influence of dissipation on momentum transport is
stronger for $d=3$ than for $d=2$. This trend is also observed in
general in the remaining transport coefficients.

One of the main limitations of our theory is its restriction to
dilute gases. In this situation, the collisional transfer
contributions to the fluxes are neglected and only their kinetic
contributions are considered. Possible extension of the present
kinetic theory to higher densities can be done in the context of
the revised Enskog theory. Preliminary results \cite{GM03} have
been focused on the uniform shear flow state to get directly the
shear viscosity coefficient. The extension of this study
\cite{GM03} to states with gradients of concentration, pressure,
and temperature is somewhat intricate due to subtleties associated
with the spatial dependence of the pair correlations functions
considered in the revised Enskog theory. On the other hand, it
must be remarked that many of the collision integrals appearing in
the Enskog description are the same as those appearing in the
Boltzmann limit so that one can take advantage of the results
reported in this paper. We plan to extend the results derived for
moderately dense mixtures of smooth {\em elastic} hard spheres
\cite{MCK83} to inelastic collisions in the near future.

\acknowledgments

Partial support of the Ministerio de Ciencia y Tecnolog\'{\i}a
(Spain) through Grant No.  FIS2004-01399 (partially financed by
FEDER funds) in the case of V.G. and ESP2003-02859 (partially
financed by FEDER funds) in the case of J.M.M. is acknowledged. V.
G. also acknowledges support from the European Community's Human
Potential Programme HPRN-CT-2002-00307 (DYGLAGEMEM).

\appendix
\section{Chapman-Enskog method}
\label{appA}

The velocity distribution function $f_1^{(1)}$ obeys the equation
\begin{equation}
\left( \partial _{t}^{(0)}+{\cal L}_{1}\right) f_{1}^{(1)}+{\cal
M} _{1}f_{2}^{(1)}=-\left( \partial _{t}^{(1)}+{\bf v}\cdot
\nabla+ {\bf g}\cdot \frac{\partial}{\partial {\bf
v}} \right)f_{1}^{(0)}\;, \label{a1}
\end{equation}
where the linear operators ${\cal L}_{1}$ and ${\cal M}_{1}$ are
defined by Eqs.\ (\ref{3.12.1}) and (\ref{3.12.2}), respectively.
A similar equation can be obtained for $f_2^{(1)}$ by
interchanging $1\leftrightarrow 2$. The action of the time
derivatives $\partial _{t}^{(1)}$ on the hydrodynamic fields is
\begin{equation}
D_{t}^{(1)}x_{1}=0,  \label{a2}
\end{equation}
\begin{equation}
D_{t}^{(1)}p=-\frac{d+2}{d}p\nabla \cdot {\bf u},  \label{a3}
\end{equation}
\begin{equation}
D_{t}^{(1)}T=-\frac{2T}{d}\nabla \cdot {\bf u},  \label{a4}
\end{equation}
\begin{equation}
D_{t}^{(1)}{\bf u}=-\rho ^{-1}{\bf \nabla }p+{\bf g},  \label{a5}
\end{equation}
where $D_{t}^{(1)}=\partial _{t}^{(1)}+{\bf u\cdot \nabla }$ and
use has been made of the results ${\bf j}_{i}^{(0)}={\bf
q}^{(0)}=\zeta ^{(1)}=0$. The last equality follows from the fact
that the cooling rate is a scalar, and corrections to first order
in the gradients can arise only from the divergence of a vector
field. However, as is demonstrated  below, there is no
contribution to the distribution function proportional to this
divergence. We note that this is special to the low density
Boltzmann equation and such terms do occur at higher densities.
\cite{GD99a} Use of Eqs.\ (\ref{a2})--(\ref{a5}) yields
\begin{eqnarray}
-\left( \partial _{t}^{(1)}+{\bf v}\cdot \nabla + {\bf g}\cdot
\frac{\partial}{\partial {\bf v}} \right)
f_{1}^{(0)} &=&-\left( \frac{\partial }{\partial
x_{1}}f_{1}^{(0)}\right) _{p,T}{\bf V} \cdot \nabla x_{1}-\left[
f_{1}^{(0)}{\bf V}+\frac{n T}{\rho}\left( \frac{
\partial }{\partial {\bf V}}f_{1}^{(0)}\right) \right] \cdot \nabla \ln p
\nonumber \\
&&+\left[ f_{1}^{(0)}+\frac{1}{2}\frac{\partial}{\partial {\bf
V}}\cdot \left({\bf V}f_{1}^{(0)}\right) \right] {\bf V}\cdot
\nabla \ln T
\nonumber \\
&&+\left( V_{k}\frac{\partial}{\partial V_{\ell
}}f_{1}^{(0)}-\frac{1}{d}\delta _{k\ell}{\bf V}\cdot
\frac{\partial}{\partial {\bf V}} f_{1}^{(0)}\right)
\nabla _{k}u_{\ell}. \label{a5bis}
\end{eqnarray}
Note that the external field does not appear in the right-hand
side of Eq.\ (\ref{a5bis}). This is due to the particular form of
the gravitational force. Using Eq.\ (\ref{a5bis}), Eq.\ (\ref{a1})
can be written as
\begin{equation}
\left( \partial _{t}^{(0)}+{\cal L}_{1}\right) f_{1}^{(1)}+{\cal
M} _{1}f_{2}^{(1)}={\bf A}_{1}\cdot \nabla x_{1}+{\bf B}_{1}\cdot
\nabla p+{\bf C}_{1}\cdot \nabla T +{\sf D}_{1}:
\nabla {\bf u}, \label{a6}
\end{equation}
where
\begin{equation}
{\bf A}_{i}({\bf V})=-\left(\frac{\partial}{\partial x_{1}}
f_{i}^{(0)}\right)_{p,T}{\bf V}, \label{a7}
\end{equation}
\begin{equation}
{\bf B}_{i}({\bf V})=-\frac{1}{p}\left[ f_{i}^{(0)}{\bf
V}+\frac{nT}{\rho } \left(\frac{\partial}{\partial {\bf
V}}f_{i}^{(0)}\right) \right] , \label{a8}
\end{equation}
\begin{equation}
{\bf C}_{i}({\bf V})=\frac{1}{T}\left[
f_{i}^{(0)}+\frac{1}{2}\frac{\partial }{\partial {\bf V}}\cdot
\left( {\bf V}f_{i}^{(0)}\right) \right] {\bf V}, \label{a9}
\end{equation}
\begin{equation}
{\sf D}_{i}({\bf V})={\bf V}\frac{\partial }{\partial {\bf V}}
f_{i}^{(0)}-\frac{1}{d}\openone {\bf V}\cdot \frac{\partial
}{\partial {\bf V}}f_{i}^{(0)}.  \label{a10}
\end{equation}
In Eqs.\ (\ref{a7})--(\ref{a10}) it is understood that $i=1,2$ and $\openone$ is the unit
tensor in $d$ dimensions. Note that the trace of ${\sf D}_{i}$ vanishes, confirming that the
distribution function does not have contribution from the divergence of
the flow field.  The solutions to Eqs.\ (\ref{a6}) are of the form
\begin{equation}
f_{i}^{(1)}={\boldsymbol {\cal A}}_{i}\cdot \nabla
x_{1}+{\boldsymbol {\cal B}}_{i}\cdot \nabla p+{\boldsymbol {\cal
C}}_{i}\cdot \nabla T+{\cal D}_{i,k\ell }\nabla _{k}u_{\ell}\;.
\label{a11}
\end{equation}
The coefficients ${\boldsymbol {\cal A}}_{i}$, ${\boldsymbol {\cal
B}}_{i}$, ${\boldsymbol {\cal C}}_{i}$, and ${\boldsymbol {\cal D}}_{i}$
are functions of the peculiar velocity ${\bf V}$ and the
hydrodynamic fields. The cooling rate depends on space through its
dependence on $x_{1}$, $p$, and $T$. The time derivative $\partial
_{t}^{(0)}$ acting on these quantities can be evaluated by the
replacement $\partial _{t}^{(0)}\rightarrow -\zeta ^{(0)}\left(
T\partial _{T}+p\partial _{p}\right) $. In addition, there are
contributions from $\partial _{t}^{(0)}$ acting on the temperature
and pressure gradients given by
\begin{eqnarray}
\partial _{t}^{(0)}\nabla T &=&-\nabla \left( T\zeta ^{(0)}\right)
=-\zeta
^{(0)}\nabla T-T\nabla \zeta ^{(0)}  \nonumber \\
&=&-\frac{\zeta ^{(0)}}{2}\nabla T -T\left[ \left( \frac{\partial
\zeta ^{(0)}}{\partial x_{1}}\right) _{p,T}\nabla x_{1}
+\frac{\zeta ^{(0)}}{p} \nabla p\right], \label{a12}
\end{eqnarray}
\begin{eqnarray}
\partial _{t}^{(0)}\nabla p &=&-\nabla \left( p\zeta ^{(0)}\right) =-\zeta
^{(0)}\nabla p-p\nabla \zeta ^{(0)}  \nonumber \\
&=&-2\zeta ^{(0)}\nabla p-p\left[ \left( \frac{\partial \zeta
^{(0)}}{
\partial x_{1}}\right) _{p,T}\nabla x_{1}-\frac{\zeta ^{(0)}}{2T}\nabla T
\right] . \label{a13}
\end{eqnarray}
The corresponding integral equations for the unknowns
${\boldsymbol {\cal A}}_{i}$, ${\boldsymbol {\cal B}}_{i}$,
${\boldsymbol {\cal C}}_{i}$, and ${\boldsymbol {\cal D}}_{i}$ are
identified as the coefficients of the independent gradients in
(\ref{a11}). This leads to Eqs.\ (\ref{3.10})--(\ref{3.12.0}).

\section{Leading Sonine approximations}
\label{appB}

In this Appendix, we get the explicit expressions of the mass,
momentum, and heat fluxes in the first Sonine approximation and
neglecting the non-Gaussian corrections to the reference
distributions $f_i^{(0)}$ (i.e., the cumulants $c_i=0$). The
procedure to get the leading order contributions in the Sonine
polynomial expansion to the transport coefficients is quite
similar to the one previously used in the three-dimensional case.
\cite{GD02} Only some partial results will be presented here.

\subsection{Leading Sonine approximation to Mass Flux}

In the case of the mass flux, the leading Sonine approximations
(lowest degree polynomial) of the quantities ${\boldsymbol {\cal
A}}_{i}$, ${\boldsymbol {\cal B}}_{i}$, ${\boldsymbol {\cal
C}}_{i}$ are
\begin{equation}
\label{b1} {\boldsymbol {\cal A}}_{1}({\bf V})\to -f_{1,M}{\bf
V}\frac{m_1m_2n}{\rho n_1T_1}D,\quad {\boldsymbol {\cal
A}}_{2}({\bf V})\to f_{2,M}{\bf V}\frac{m_1m_2n}{\rho n_2T_2}D
\end{equation}
\begin{equation}
\label{b2} {\boldsymbol {\cal B}}_{1}({\bf V})\to -f_{1,M}{\bf
V}\frac{\rho}{p n_1T_1}D_p ,\quad {\boldsymbol {\cal B}}_{2}({\bf
V})\to f_{2,M}{\bf V}\frac{\rho}{p n_2T_2}D_p
\end{equation}
\begin{equation}
\label{b3} {\boldsymbol {\cal C}}_{1}({\bf V})\to -f_{1,M}{\bf
V}\frac{\rho}{T n_1T_1}D',\quad {\boldsymbol {\cal C}}_{2}({\bf
V})\to f_{2,M}{\bf V}\frac{\rho}{T n_2T_2}D',
\end{equation}
where $f_{i,M}$ are the Maxwellian distributions (\ref{2.18}).
Multiplication of Eqs.\ (\ref{3.10})--(\ref{3.12}) by $m_1 {\bf
V}$ and integrating over the velocity yields
\begin{equation}
\left[ -\zeta ^{(0)}\left( T\partial _{T}+p\partial _{p}\right)
+\nu \right] \left(-\frac{m_1m_2n}{\rho}D\right)=-\left(\frac{
\partial }{\partial x_{1}}n_{1}T_{1}\right)_{p,T}
-\rho\left( \frac{\partial \zeta ^{(0)}}{\partial x_{1}}\right)
_{p,T}\left(D_p+D'\right) ,  \label{b4}
\end{equation}
\begin{equation}
\left[ -\zeta ^{(0)}\left( T\partial _{T}+p\partial _{p}\right)
-2\zeta ^{(0)}+\nu \right]\left(-\frac{\rho}{p}D_p\right) =-\frac{
n_{1}T_{1}}{p}\left( 1-\frac{m_{1}nT}{\rho T_{1}}\right)
-\frac{\rho\zeta ^{(0)}}{p}D',  \label{b5}
\end{equation}
\begin{equation}
\left[-\zeta ^{(0)}\left( T\partial _{T}+p\partial _{p}\right)
-\frac{1}{2} \zeta ^{(0)}+\nu \right]
\left(-\frac{\rho}{T}D'\right)= \frac{\rho\zeta ^{(0)}}{2T}D_p.
\label{b6}
\end{equation}
Here, $\nu $ is the collision frequency defined by
\begin{eqnarray}
\nu  &=&\frac{1}{dn_{1}T_{1}}\int d{\bf V}_{1}m_{1}{\bf
V}_{1}\cdot \left[ {\cal L}_{1}(f_{1,M}{\bf V}_{1})-\delta \gamma
{\cal M}_{1}(f_{2,M}{\bf V}_{2})
\right]   \nonumber \\
&=&-\frac{1}{dn_{1}T_{1}}\int d{\bf V}_{1}m_{1}{\bf V}_{1}\cdot
\left( J_{12}[{\bf v}_{1}|f_{1,M}{\bf V}_{1},f_{2}^{(0)}]-\delta
\gamma J_{12}[{\bf v}_{1}|f_{1}^{(0)},f_{2,M}{\bf V}_{2}]\right),
\label{b7}
\end{eqnarray}
where $\delta\equiv x_1/x_2$. The evaluation of the collision integral
(\ref{b7}) is made in Appendix \ref{appC}. The self-collision
terms of ${\cal L}_{i}$ arising from $J_{11}$ do not occur in Eq.\
(\ref{b7}) since they conserve momentum for species $1$. From
dimensional analysis, $D\sim T^{1/2}$, $ D_p\sim p T^{-1/2}$, and
$D'\sim p T^{-1/2}$ so the temperature and pressure derivatives
can be performed in Eqs.\ (\ref{b4})--(\ref{b6}). After performing
them, one gets the expressions (\ref{3.13}), (\ref{3.14}), and
(\ref{3.15}) for the (reduced) coefficients $D^*$, $D_p^*$, and
$D^{\prime\ast}$, respectively.

\subsection{Leading Sonine approximation to Pressure Tensor}

In the case of the pressure tensor, the leading Sonine
approximation for the function ${\cal D}_{i,k\ell}$ is
\begin{equation}
\label{b8} {\cal D}_{i,k\ell}({\bf V})\to -f_{i,M}({\bf V}) \frac{\eta_{i}}{T}
R_{i,k\ell}({\bf V}),\quad i=1,2
\end{equation}
where
\begin{equation}
\label{b9} R_{i,k\ell}({\bf V})=m_i\left( V_{k}V_{\ell}-
\frac{1}{d}V^2\delta_{k\ell}\right),
\end{equation}
and
\begin{equation}
\label{b10} \eta_i=-\frac{1}{(d-1)(d+2)}\frac{T}{n_iT_i^2}\int
d{\bf v} R_{i,k\ell}({\bf V}){\cal D}_{i,k\ell}({\bf V}).
\end{equation}
The shear viscosity $\eta$ in this approximation can be written as
\begin{equation}
\label{b11} \eta=\frac{p}{\nu_0}\left(x_1\gamma_1^2
\eta_1^*+x_2\gamma_2^2 \eta_2^*\right),
\end{equation}
where $\eta_i^*=\nu_0\eta_i$. The integral equations for the
(reduced) coefficients $\eta_i^*$ are decoupled from the remaining
transport coefficients. The two coefficients $\eta_{i}^*$ are
obtained by multiplying Eqs.\ (\ref{3.12.0}) with $R_{i,k\ell}$
and integrating over the velocity to get the coupled set of
equations
\begin{equation}
\label{b12} \left(
\begin{array}{cc}
\tau_{11}-\frac{1}{2}\zeta^{*}& \tau_{12}\\
\tau_{21}&\tau_{22}-\frac{1}{2}\zeta^{*}
\end{array}
\right) \left(
\begin{array}{c}
\eta_{1}^*\\
\eta_{2}^*
\end{array}
\right) =\left(
\begin{array}{c}
\gamma_1^{-1}\\
\gamma_2^{-1}
\end{array}
\right).
\end{equation}
The (reduced) collision frequencies $\tau_{ij}$ are given in terms
of the linear collision operator by
\begin{equation}
\label{b13}
\tau_{ii}=\frac{1}{(d-1)(d+2)}\frac{1}{n_iT_i^2\nu_0}\int d{\bf
v}_1 R_{i,k\ell} {\cal L}_i\left(f_{i,M}R_{i,k\ell}\right),
\end{equation}
\begin{equation}
\label{b14}
\tau_{ij}=\frac{1}{(d-1)(d+2)}\frac{1}{n_iT_i^2\nu_0}\int d{\bf
v}_1 R_{i,k\ell} {\cal M}_i\left(f_{j,M}R_{j,k\ell}\right) , \quad
i\neq j.
\end{equation}
The evaluation of these collision integrals is also given in
Appendix \ref{appC}. The solution of Eq.\ (\ref{b12})
is elementary and yields Eq.\ (\ref{3.18}).

\subsection{Leading Sonine approximation to Heat Flux}

The heat flux requires going up to the second Sonine
approximation. In this case, the quantities ${\boldsymbol {\cal
A}}_{i}$, ${\boldsymbol {\cal B}}_{i}$, ${\boldsymbol {\cal
C}}_{i}$ are taken to be
\begin{equation}
\label{b15} {\boldsymbol {\cal A}}_{1}({\bf V})\to
f_{1,M}\left[-\frac{m_1m_2n}{\rho n_1T_1}D{\bf V}+d_1''{\bf
S}_1({\bf V}) \right] ,\quad {\boldsymbol {\cal A}}_{2}({\bf
V})\to f_{2,M}\left[\frac{m_1m_2n}{\rho n_2T_2}D{\bf V}+d_2''{\bf
S}_2({\bf V})\right]
\end{equation}
\begin{equation}
\label{b16} {\boldsymbol {\cal B}}_{1}({\bf V})\to
f_{1,M}\left[-\frac{\rho}{p n_1T_1}D_p{\bf V}+\ell_1{\bf S}_1({\bf
V}) \right] ,\quad {\boldsymbol {\cal B}}_{2}({\bf V})\to
f_{2,M}\left[\frac{\rho}{p n_2T_2}D_p{\bf V}+\ell_2{\bf S}_2({\bf
V})\right]
\end{equation}
\begin{equation}
\label{b17} {\boldsymbol {\cal C}}_{1}({\bf V})\to
f_{1,M}\left[-\frac{\rho}{T n_1T_1}D'{\bf V}+\lambda_1{\bf
S}_1({\bf V}) \right] ,\quad {\boldsymbol {\cal C}}_{2}({\bf
V})\to f_{2,M}\left[\frac{\rho}{T n_2T_2}D'{\bf V}+\lambda_2{\bf
S}_2({\bf V})\right],
\end{equation}
where
\begin{equation}
\label{b18} {\bf S}_i({\bf
V})=\left(\frac{1}{2}m_iV^2-\frac{d+2}{2}T_i\right){\bf V}.
\end{equation}
In these equations, it is understood that $D$, $D_p$ and $D'$ are
given by Eqs.\ (\ref{3.13}), (\ref{3.14}), and (\ref{3.15}),
respectively. The coefficients $d_i''$, $\ell_i$ and $\lambda_i$
are defined as
\begin{equation}
\left(
\begin{array}{c}
d_i'' \\
\ell_i \\
\lambda_i
\end{array}
\right) =\frac{2}{d(d+2)}\frac{m_i}{n_iT_i^3} \int d{\bf v}\,{\bf
S}_i({\bf V})\cdot\left(
\begin{array}{c}
{\boldsymbol {\cal A}}_{i} \\
{\boldsymbol {\cal B}}_{i} \\
{\boldsymbol {\cal C}}_{i}
\end{array}
\right). \label{b19}
\end{equation}
These coefficients can be determined by multiplying Eqs.\
(\ref{3.10})--(\ref{3.12}) (and their counterparts for the species
2) by ${\bf S}_i({\bf V})$ and integrating over the velocity. The
final expressions can be obtained by taking into account that
$d_1''\sim T^{-3/2}$, $\ell_1\sim T^{-3/2}/p$, and $\lambda_1\sim
T^{-5/2}$ and the results
\begin{equation}
\int d{\bf v}\,m_{1}{\bf S}_1({\bf V})\cdot {\bf A}_{1}=-
\frac{d(d+2)}{4}\frac{n_1T^2}{m_1} \left( \frac{
\partial }{\partial x_{1}}\gamma_1^2\right) _{p,T},  \label{b20}
\end{equation}
\begin{equation}
\int d{\bf v}\,m_{1}{\bf S}_{1}({\bf V})\cdot {\bf B}_{1}=0,
\label{b21}
\end{equation}
\begin{equation}
\int d{\bf v}\,m_{1}{\bf S}_{1}({\bf V})\cdot {\bf
C}_{1}=-\frac{d(d+2)}{2} \frac{n_1T_1^2}{m_1T}. \label{b22}
\end{equation}

By using matrix notation, the coupled set of six equations for the
quantities
\begin{equation}
\label{4.10} \{d_1^*, d_2^*, \ell_1^*, \ell_2^*, \lambda_1^*,
\lambda_2^*\}
\end{equation}
can be written as
\begin{equation}
\label{4.11} \Lambda_{\sigma \sigma'}X_{\sigma'}=Y_{\sigma}.
\end{equation}
Here, $d_i^*\equiv T\nu_0 d_i^{\prime\prime}$, $\ell_i^*\equiv
pT\nu_0 \ell_i$, $\lambda_i^*\equiv T^2\nu_0 \lambda_i$,
$X_{\sigma'}$ is the column matrix defined by the set (\ref{4.10})
and $\Lambda_{\sigma \sigma'}$ is the square matrix
\begin{equation}
\label{4.12} \Lambda=\left(
\begin{array} {cccccc}
\nu_{11}-\frac{3}{2}\zeta^{*}& \nu_{12}& -\left(\frac{\partial
\zeta ^{*}}{\partial x_{1}}\right)_{p,T}&0&
-\left( \frac{\partial \zeta ^{*}}{\partial x_{1}}\right)_{p,T}&0 \\
\nu_{21}&\nu_{22}-\frac{3}{2}\zeta^{*}&0& -\left( \frac{\partial
\zeta ^{*}}{\partial x_{1}}\right)_{p,T}&0&
-\left( \frac{\partial \zeta ^{*}}{\partial x_{1}}\right)_{p,T}\\
0& 0& \nu_{11}-\frac{5}{2}\zeta^{*}& \nu_{12}&
-\zeta^{*}&0\\
0& 0 & \nu_{21} & \nu_{22}-\frac{5}{2}\zeta^{*} & 0 &
-\zeta^{*}\\
0& 0& \zeta^{*}/2&0&\nu_{11}-\zeta^{*}& \nu_{12}\\
0& 0& 0&\zeta^{*}/2&\nu_{21}&\nu_{22}-\zeta^{*}
\end{array}
\right).
\end{equation}
The column matrix ${\bf Y}$ is
\begin{equation}
\label{4.13} {\bf Y}=\left(
\begin{array}{c}
Y_1\\
Y_2\\
Y_3\\
Y_4\\
Y_5\\
Y_6
\end{array}
\right),
\end{equation}
where \cite{note}
\begin{equation}
\label{4.14} Y_1= \frac{D^*}{ x_1\gamma_1^2}\left(\omega_{12}-\zeta^{*}\right)-\frac{1}{ \gamma_1^2}
\left(\frac{\partial \gamma_1}{\partial x_1} \right)_{p,T},\quad Y_2= -\frac{D^*}{
x_2\gamma_2^2}\left(\omega_{21}-\zeta^{*}\right)-\frac{1}{ \gamma_2^2} \left(\frac{\partial \gamma_2}{\partial
x_1} \right)_{p,T},
\end{equation}
\begin{equation}
\label{4.15} Y_3=\frac{
D_p^*}{x_1\gamma_1^2}\left(\omega_{12}-\zeta^{*}\right),\quad
Y_4=-\frac{
D_p^*}{x_2\gamma_2^2}\left(\omega_{21}-\zeta^{*}\right),
\end{equation}
\begin{equation}
\label{4.16} Y_5= -\frac{1}{\gamma_1}+\frac{D^{\prime\ast}}{
x_1\gamma_1^2}\left(\omega_{12}-\zeta^{*}\right),\quad Y_6=
-\frac{1}{\gamma_2}-\frac{D^{\prime\ast}}{
x_2\gamma_2^2}\left(\omega_{21}-\zeta^{*}\right).
\end{equation}
Here, we have introduced the (reduced) collision frequencies
\begin{equation}
\label{4.17}
\nu_{ii}=\frac{2}{d(d+2)}\frac{m_i}{n_iT_i^3\nu_0}\int d{\bf v}_1
{\bf S}_i \cdot {\cal L}_i\left(f_{i,M}{\bf S}_i\right),
\end{equation}
\begin{equation}
\label{4.18}
\nu_{ij}=\frac{2}{d(d+2)}\frac{m_i}{n_iT_i^3\nu_0}\int d{\bf v}_1
{\bf S}_i \cdot {\cal M}_i\left(f_{j,M}{\bf S}_j\right), \quad
i\neq j,
\end{equation}
\begin{equation}
\label{4.19} \omega_{12}=\frac{2}{d(d+2)}\frac{m_1}{n_1T_1^2\nu_0}
\left[\int d{\bf v}_1 {\bf S}_1\cdot {\cal L}_1(f_{1,M}{\bf
V}_1)-\delta \gamma \int d{\bf v}_1 {\bf S}_1\cdot {\cal
M}_1(f_{2,M}{\bf V}_2)\right],
\end {equation}
\begin{equation}
\label{4.19.1}
\omega_{21}=\frac{2}{d(d+2)}\frac{m_2}{n_2T_2^2\nu_0} \left[\int
d{\bf v}_1 {\bf S}_2\cdot {\cal L}_2(f_{2,M}{\bf
V}_1)-\frac{1}{\delta \gamma} \int d{\bf v}_1 {\bf S}_2\cdot {\cal
M}_2(f_{1,M}{\bf V}_2)\right].
\end {equation}
The expressions of the collision integrals (\ref{4.17}),
(\ref{4.18}), and (\ref{4.19}) are given in Appendix \ref{appC}.
The solution to Eq.\ (\ref{4.11}) is
\begin{equation}
\label{4.20} X_{\sigma}=\left(\Lambda^{-1}\right)_{\sigma
\sigma'}Y_{\sigma'}.
\end{equation}
From this relation one gets the expressions (\ref{4.n2}),
(\ref{4.n3}), and (\ref{4.n4}) for the coefficients $d_i^*$,
$\ell_i^*$ and $\lambda_i^*$, respectively.

\section{Collision integrals}
\label{appC}

In this Appendix we compute the different collision integrals
appearing in the expressions of the transport coefficients. To
simplify all the integrals, we use the property
\begin{equation}
\int d{\bf v}_{1}h({\bf V}_{1})J_{ij}\left[ {\bf
v}_{1}|f_{i},f_{j}\right] =\sigma _{ij}^{d-1}\int \,d{\bf
v}_{1}\,\int \,d{\bf v}_{2}f_{i}( {\bf V}_{1})f_{j}({\bf V}_{2})
\int d\widehat{\boldsymbol {\sigma }}\,\Theta
(\widehat{\boldsymbol {\sigma}} \cdot {\bf
g}_{12})(\widehat{\boldsymbol {\sigma }}\cdot {\bf
g}_{12})\,\left[ h( {\bf V}_{1}^{^{\prime \prime }})-h({\bf
V}_{1})\right] \;,  \label{c1}
\end{equation}
with
\begin{equation}
{\bf V}_{1}^{^{\prime \prime }}={\bf V}_{1}-\mu _{ji}(1+\alpha
_{ij})( \widehat{\boldsymbol {\sigma }}\cdot {\bf
g}_{12})\widehat{\boldsymbol {\sigma}}\;. \label{c2}
\end{equation}
This result applies for both $i=j$ and $i\neq j$.

Let us start with the collision frequency $\nu$ defined by Eq.\
(\ref{b7}). Use of the identity (\ref{c2}) in (\ref{b7}) gives
\begin{eqnarray}
\nu &=&\frac{m_1}{dn_1T_1}B_3\sigma _{12}^{d-1}\mu _{21}(1+\alpha
_{12}) \int \,d{\bf V}_{1}\,\int \,d{\bf V}_{2}\,g_{12}\left[
f_{1,M}( {\bf V}_{1})f_{2}^{(0)}({\bf V}_{2})({\bf V}_{1}\cdot
{\bf g}_{12})\right.
\nonumber \\
&&\left. -\delta \gamma f_{1}^{(0)}({\bf V}_{1})f_{2,M}({\bf
V}_{2})({\bf V} _{2}\cdot {\bf g}_{12})\right],  \label{c3}
\end{eqnarray}
where use has been made of the result
\begin{equation}
\label{c4} \int d\widehat{\boldsymbol{\sigma}}\, \Theta
(\widehat{{\boldsymbol {\sigma }}} \cdot {\bf g}_{12})\,
(\widehat{\boldsymbol{\sigma}}\cdot {\bf g}_{12})^k
\widehat{\boldsymbol{\sigma}}=B_{k+1} g_{12}^{k-1}{\bf g}_{12},
\end{equation}
whith \cite{EB02}
\begin{equation}
\label{c5} B_k\equiv \int d\widehat{\boldsymbol{\sigma}}\, \Theta
(\widehat{{\boldsymbol {\sigma }}} \cdot {\bf g}_{12})\,
(\widehat{\boldsymbol{\sigma}}\cdot {\widehat{\bf
g}}_{12})^k=\pi^{(d-1)/2} \frac{
\Gamma\left(\frac{k+1}{2}\right)}{\Gamma\left(\frac{k+d}{2}\right)}.
\end{equation}
Substitution of the Maxwellian approximation (\ref{2.18}) for
$f_i^{(0)}$ gives
\begin{equation}
\nu =\frac{2}{d}\frac{\pi^{(d-1)/2}}
{\Gamma\left(\frac{d+3}{2}\right)} \nu_0 (1+\alpha _{12}) \pi^{-d}
\left(\theta_1\theta_2\right)^{d/2} \int \,d{\bf c}_{1}\,\int
\,d{\bf c}_{2}\,y\, e^{-(\theta
_{1}c_{1}^{2}+\theta_{2}c_{2}^{2})}\left[ x_2\gamma_1^{-1}({\bf
c}_{1}\cdot {\bf y}) -x_1\gamma_2^{-1}({\bf c}_{2}\cdot {\bf
y})\right], \label{c6}
\end{equation}
where ${\bf c}_{i}\equiv {\bf V}_{i}/v_{0}$ and ${\bf y}\equiv {\bf c}_1-{\bf
c}_2$. The integral can be performed by the change of variables
$\{{\bf c}_1,{\bf c}_2\}\to \{{\bf y},{\bf z}\}$, where ${\bf
z}\equiv \theta _{1} {\bf c}_{1}+\theta _{2}{\bf c}_{2}$ and the
Jacobian is $\left( \theta _{1}+\theta _{2}\right) ^{-d}$. With
this change, Eq.\ (\ref{c6}) becomes
\begin{equation}
\label{c6.1} \nu=\frac{2}{d}\frac{\pi^{(d-1)/2}}
{\Gamma\left(\frac{d+3}{2}\right)} \nu_0 (1+\alpha _{12}) \pi^{-d}
\left(\theta_1\theta_2\right)^{(d+1)/2}\left( \theta _{1}+\theta
_{2}\right) ^{-(1+d)}(x_2\mu_{21}+x_1\mu_{12})\int \,d{\bf
y}\,\int \,d{\bf z}\,y^3 e^{-(a y^{2}+b z^{2})},
\end{equation}
where $a\equiv \theta_1\theta_2\left( \theta _{1}+\theta
_{2}\right) ^{-1}$ and $b\equiv \left( \theta _{1}+\theta
_{2}\right) ^{-1}$. The integral (\ref{c6.1}) can be easily
computed and one directly gets the result (\ref{3.16}) given in
the text for the reduced collision frequency $\nu^*=\nu/\nu_0$.

The collision frequencies $\tau_{ij}$ defined by Eqs.\
(\ref{4.17}) and (\ref{4.18}) involve collision integrals of the
form
\begin{equation}
\label{c8} \int d{\bf v}_1 {\bf V}_{1}{\bf V}_{1}
 J_{ij}[f_i,f_j]=\sigma_{ij}^{d-1}\int\, d{\bf
v}_1\int\, d{\bf v}_2\, f_i({\bf V}_1) f_j({\bf V}_2)  \int
d\widehat{\boldsymbol {\sigma}}\,\Theta (\widehat{\boldsymbol
{\sigma}} \cdot {\bf g}_{12})(\widehat{\boldsymbol {\sigma}}\cdot
{\bf g}_{12})\,\left[ {\bf V}_1''{\bf V}_{1}''-{\bf V}_{1}{\bf
V}_{1}\right],
\end{equation}
where the identity (\ref{c1}) has been used. The scattering rule
(\ref{c2}) gives
\begin{equation}
\label{c9} {\bf V}_{1}''{\bf V}_{1}''-{\bf V}_{1}{\bf
V}_{1}=-\mu_{ji}(1+ \alpha_{ij})(\widehat{\boldsymbol
{\sigma}}\cdot {\bf g}_{12})\left[ {\bf
G}_{ij}\widehat{\boldsymbol
{\sigma}}+\widehat{\boldsymbol{\sigma}}{\bf G}_{ij} +\mu_{ji}({\bf
g}_{12}\widehat{\boldsymbol{\sigma}}+\widehat{\boldsymbol{\sigma}}{\bf
g}_{12}) -\mu_{ji}(1+\alpha_{ij})(\widehat{\boldsymbol
{\sigma}}\cdot {\bf g}_{12})
\widehat{\boldsymbol{\sigma}}\widehat{\boldsymbol{\sigma}}\right],
\end{equation}
where ${\bf G}_{ij}=\mu_{ij}{\bf V}_1+\mu_{ji}{\bf V}_2$.
Substitution of Eq.\ (\ref{c9}) into Eq.\ (\ref{c8}) allows the
angular integral to be performed with the result
\begin{eqnarray}
\label{c10} \int d\widehat{\boldsymbol {\sigma}}\,\Theta
(\widehat{\boldsymbol {\sigma}}\cdot {\bf g
}_{12})(\widehat{\boldsymbol {\sigma}}\cdot {\bf g}_{12})&&\left[
{\bf V}_{1}''{\bf V}_{1}''-{\bf V}_{1}{\bf V}_{1}\right]= -B_3 m_i
\mu_{ji}(1+\alpha _{ij})\left[g_{12}({\bf G}_{ij}{\bf g}_{12}+
{\bf g}_{12}{\bf G}_{ij})\right. \nonumber\\
& &\left. +3\frac{\mu_{ji}}{d+3}
(1+\frac{2d}{3}-\alpha_{ij})g_{12}{\bf g}_{12}{\bf g}_{12}
-\frac{\mu_{ji}}{d+3}(1+\alpha_{ij})g^3\openone\right].
\end{eqnarray}
Using (\ref{c10}) the integrals defining $\tau_{ij}$ can be calculated by the same
mathematical steps as those made before for $\nu$. After a lengthy calculation, one gets
\begin{eqnarray}
\label{c11} \int d{\bf v}_1 R_{1,k\ell}J_{12}[f_1^{(0)},
f_{2,M}R_{2,k\ell}] &=&
-\frac{\pi^{(d-1)/2}}{2d\Gamma\left(\frac{d}{2}\right)}m_1
m_2n_1n_2\mu_{21}(1+\alpha_{12}) \sigma_{12}^{d-1}
v_0^5 \left(\theta_1\theta_2\right)^{-1/2}\nonumber\\
& \times &
\left\{2(d+3)(d-1)(\mu_{12}\theta_2-\mu_{21}\theta_1)\theta_2^{-2}
\left(\theta_1+\theta_2\right)^{-1/2}\right.\nonumber\\
& & +3(d-1)\mu_{21}\left(1+\frac{2d}{3}-\alpha_{12}\right)
\theta_2^{-2}\left(\theta_1+\theta_2\right)^{1/2}\nonumber\\
& & \left.
-\left[2d(d+1)-4\right]\theta_2^{-1}\left(\theta_1+\theta_2\right)^{-1/2}\right\},
\end{eqnarray}
\begin{eqnarray}
\label{c12} \int d{\bf v}_1 R_{1,k\ell}J_{12}[f_{1,M}R_{1,k\ell},
f_2^{(0)}] &=&
-\frac{\pi^{(d-1)/2}}{2d\Gamma\left(\frac{d}{2}\right)}m_1
m_2n_1n_2\mu_{21}(1+\alpha_{12}) \sigma_{12}^{d-1}
v_0^5 \left(\theta_1\theta_2\right)^{-1/2}\nonumber\\
& \times &
\left\{2(d+3)(d-1)(\mu_{12}\theta_2-\mu_{21}\theta_1)\theta_1^{-2}
\left(\theta_1+\theta_2\right)^{-1/2}\right.\nonumber\\
& & +3(d-1)\mu_{21}\left(1+\frac{2d}{3}-\alpha_{12}\right)
\theta_1^{-2}\left(\theta_1+\theta_2\right)^{1/2}\nonumber\\
& & \left.
+\left[2d(d+1)-4\right]\theta_1^{-1}\left(\theta_1+\theta_2\right)^{-1/2}\right\},
\end{eqnarray}
\begin{eqnarray}
\label{c13} \int d{\bf v}_1
R_{1,k\ell}\left\{J_{11}[f_1^{(0)},f_{1,M}R_{1,k\ell}]
+J_{11}[f_{1,M}R_{1,k\ell},f_1^{(0)}]\right\}
&=&-\frac{\pi^{(d-1)/2}}{\Gamma\left(\frac{d}{2}\right)}m_1^2
n_1^2(1+\alpha_{11})\sigma_1^{d-1} (T_1/m_1)^{5/2}\nonumber\\
& & \times\frac{6(d-1)}{d}
\left(1+\frac{2d}{3}-\alpha_{11}\right).
\end{eqnarray}
The corresponding expressions for $\tau_{ij}$ can be easily
inferred from Eqs. \ (\ref{c11})--(\ref{c13}).

The collision frequencies $\nu_{ij}$ and $\omega_{ij}$ that
determine the heat flux are defined by Eqs.\ (\ref{4.17}),
(\ref{4.18}), and (\ref{4.19}), respectively. To evaluate these
collision integrals, one needs the partial results
\begin{eqnarray}
\label{c15} {\bf S}_{i}({\bf V}_{1}^{^{\prime \prime}})-{\bf
S}_{i}({\bf V}_{1}) &=& \frac{m_{i}}{2}(1+\alpha
_{ij})\mu_{ji}(\widehat{\boldsymbol {\sigma}}\cdot {\bf g
}_{12})\left\{ \left[
(1-\alpha_{ij}^{2})\mu_{ji}^{2}(\widehat{\boldsymbol
{\sigma}}\cdot {\bf
g}_{12})^{2}-G_{ij}^{2}-\mu_{ji}^{2}g_{12}^{2}\right.
\right.   \nonumber \\
&&\left. -2\mu_{ji}({\bf g}_{12}\cdot {\bf
G}_{ij})+2(1+\alpha_{ij})\mu _{ji}(\widehat{\boldsymbol
{\sigma}}\cdot {\bf g}_{12})(\widehat{\boldsymbol {\sigma }} \cdot
{\bf G}_{ij})+(d+2)\frac{T_{i}}{m_{i}}\right] \widehat{\boldsymbol
{\sigma}}
\nonumber \\
&&-\left[ (1-\alpha_{ij})\mu_{ji}(\widehat{\boldsymbol {\sigma
}}\cdot {\bf g}_{12} )+2(\widehat{\boldsymbol {\sigma}}\cdot {\bf
G}_{ij})\right] {\bf G}_{ij}
\nonumber \\
&&\left.
-\mu_{ji}\left[(1-\alpha_{ij})\mu_{ji}(\widehat{\boldsymbol
{\sigma}} \cdot {\bf g}_{12})+2(\widehat{\boldsymbol{\sigma}}\cdot
{\bf G}_{ij}) \right] {\bf g}_{12}\right\},
\end{eqnarray}
\begin{eqnarray}
\label{c16} \int d\widehat{\boldsymbol {\sigma}}\,\Theta
(\widehat{\boldsymbol {\sigma}}\cdot {\bf g
}_{12})(\widehat{\boldsymbol {\sigma}}\cdot {\bf g}_{12}) &&\left[
{\bf S}_{i}( {\bf V}_{1}^{^{\prime \prime }})-{\bf S}_{i}({\bf
V}_{1})\right] =-\frac{
m_{i}}{2}\frac{\pi^{(d-1)/2}}{\Gamma\left(\frac{d+3}{2}\right)}
(1+\alpha_{ij})\mu_{ji}\nonumber\\
& &\times \left\{ \left[
g_{12}G_{ij}^{2}+\mu_{ji}^{2}\frac{4\alpha_{ij}^{2}-(d+3)\alpha
_{ij}+2(d+1)}{d+3} g_{12}^{3}\right.\right.\nonumber\\
& & \left. -2\mu_{ji}\frac{3\alpha_{ij}-2d-3}{d+3} g_{12}\left(
{\bf g}_{12}\cdot {\bf G}_{ij}\right) -(d+2)\frac{T_{i}}{m_{i}}
g_{12}\right] {\bf g}_{12}  \nonumber \\
&&\left. +\left[ 2 g_{12}\left( {\bf g}_{12}\cdot {\bf
G}_{ij}\right) -\mu_{ji}\frac{(d+5)\alpha_{ij}-d-1}{d+3}
g_{12}^{3}\right] {\bf G}_{ij}\right\} \;.
\end{eqnarray}
The integrals $\omega_{ij}$ and $\nu_{ij}$ can be explicitly
evaluated by using (\ref{c16}) and the same mathematical steps as
before. After a lengthy algebra, one gets
\begin{eqnarray}
\label{c17} \omega_{12}&=&\frac{\pi^{(d-1)/2}}
{\Gamma\left(\frac{d}{2}\right)}\frac{2}{d\sqrt{2}}
\left(\frac{\sigma_1}{\sigma_{12}}\right)^{d-1}x_1
\theta_1^{-1/2}(1-\alpha_{11}^2)\nonumber\\
& & +\frac{\pi^{(d-1)/2}}
{\Gamma\left(\frac{d}{2}\right)}\frac{2}{d(d+2)}
x_1\mu_{21}(1+\alpha_{12})(\theta_1+\theta_2)^{-1/2}\theta_1^{1/2}
\theta_2^{-3/2}\left(\frac{x_2}{x_1}A-\gamma B\right),
\end{eqnarray}
\begin{eqnarray}
\label{c18}
\nu_{11}&=&\frac{\pi^{(d-1)/2}}{\Gamma\left(\frac{d}{2}\right)}
\frac{8}{d(d+2)}\left(\frac{\sigma_1}{\sigma_{12}}\right)^{d-1}
x_1 (2\theta_1)^{-1/2}
(1+\alpha_{11})\left[\frac{d-1}{2}+\frac{3}{16}(d+8)(1-\alpha_{11})\right]\nonumber\\
& & +\frac{\pi^{(d-1)/2}}{\Gamma\left(\frac{d}{2}\right)}
\frac{1}{d(d+2)}x_2\mu_{21}(1+\alpha_{12})\left(\frac{\theta_1}
{\theta_2(\theta_1+\theta_2)}\right)^{3/2}\left[E-(d+2)\frac{\theta_1+\theta_2}{\theta_1}A\right],
\end{eqnarray}
\begin{equation}
\label{c19}
\nu_{12}=-\frac{\pi^{(d-1)/2}}{\Gamma\left(\frac{d}{2}\right)}
\frac{1}{d(d+2)}x_2\frac{\mu_{21}^2}{\mu_{12}}(1+\alpha_{12})\left(\frac{\theta_1}
{\theta_2(\theta_1+\theta_2)}\right)^{3/2}\left[F+(d+2)\frac{\theta_1+\theta_2}{\theta_2}B\right].
\end{equation}
In the above equations we have introduced the quantities
\cite{note}
\begin{eqnarray}
\label{c20}
A&=&(d+2)(2\beta_{12}+\theta_2)+\mu_{21}(\theta_1+\theta_2)\left\{(d+2)(1-\alpha_{12})
-[(11+d)\alpha_{12}-5d-7]\beta_{12}\theta_1^{-1}\right\}\nonumber\\
& &
+3(d+3)\beta_{12}^2\theta_1^{-1}+2\mu_{21}^2\left(2\alpha_{12}^{2}-\frac{d+3}{2}\alpha
_{12}+d+1\right)\theta_1^{-1}(\theta_1+\theta_2)^2-
(d+2)\theta_2\theta_1^{-1}(\theta_1+\theta_2), \nonumber\\
\end{eqnarray}
\begin{eqnarray}
\label{c21} B&=&
(d+2)(2\beta_{12}-\theta_1)+\mu_{21}(\theta_1+\theta_2)\left\{(d+2)(1-\alpha_{12})
+[(11+d)\alpha_{12}-5d-7]\beta_{12}\theta_2^{-1}\right\}\nonumber\\
& &
-3(d+3)\beta_{12}^2\theta_2^{-1}-2\mu_{21}^2\left(2\alpha_{12}^{2}-\frac{d+3}{2}\alpha
_{12}+d+1\right)\theta_2^{-1}(\theta_1+\theta_2)^2+
(d+2)(\theta_1+\theta_2), \nonumber\\
\end{eqnarray}
\begin{eqnarray}
\label{c22} E&=&
 2\mu_{21}^2\theta_1^{-2}(\theta_1+\theta_2)^2
\left(2\alpha_{12}^{2}-\frac{d+3}{2}\alpha_{12}+d+1\right)
\left[(d+2)\theta_1+(d+5)\theta_2\right]\nonumber\\
& & -\mu_{21}(\theta_1+\theta_2)
\left\{\beta_{12}\theta_1^{-2}[(d+2)\theta_1+(d+5)\theta_2][(11+d)\alpha_{12}
-5d-7]\right.\nonumber\\
& & \left.
-\theta_2\theta_1^{-1}[20+d(15-7\alpha_{12})+d^2(1-\alpha_{12})-28\alpha_{12}]
-(d+2)^2(1-\alpha_{12})\right\}
\nonumber\\
& & +3(d+3)\beta_{12}^2\theta_1^{-2}[(d+2)\theta_1+(d+5)\theta_2]+
2\beta_{12}\theta_1^{-1}[(d+2)^2\theta_1+(24+11d+d^2)\theta_2]
\nonumber\\
& & +(d+2)\theta_2\theta_1^{-1}
[(d+8)\theta_1+(d+3)\theta_2]-(d+2)(\theta_1+\theta_2)\theta_1^{-2}\theta_2
[(d+2)\theta_1+(d+3)\theta_2],\nonumber\\
\end{eqnarray}
\begin{eqnarray}
\label{c23} F&=&
 2\mu_{21}^2\theta_2^{-2}(\theta_1+\theta_2)^2
\left(2\alpha_{12}^{2}-\frac{d+3}{2}\alpha_{12}+d+1\right)
\left[(d+5)\theta_1+(d+2)\theta_2\right]\nonumber\\
& & -\mu_{21}(\theta_1+\theta_2)
\left\{\beta_{12}\theta_2^{-2}[(d+5)\theta_1+(d+2)\theta_2][(11+d)\alpha_{12}
-5d-7]\right.\nonumber\\
& & \left.
+\theta_1\theta_2^{-1}[20+d(15-7\alpha_{12})+d^2(1-\alpha_{12})-28\alpha_{12}]
+(d+2)^2(1-\alpha_{12})\right\}
\nonumber\\
& & +3(d+3)\beta_{12}^2\theta_2^{-2}[(d+5)\theta_1+(d+2)\theta_2]-
2\beta_{12}\theta_2^{-1}[(24+11d+d^2)\theta_1+(d+2)^2\theta_2]
\nonumber\\
& & +(d+2)\theta_1\theta_2^{-1}
[(d+3)\theta_1+(d+8)\theta_2]-(d+2)(\theta_1+\theta_2)\theta_2^{-1}
[(d+3)\theta_1+(d+2)\theta_2]. \nonumber\\
\end{eqnarray}
Here, $\beta_{12}=\mu_{12}\theta_2-\mu_{21}\theta_1$. From Eqs.\
(\ref{c17})--(\ref{c23}), one easily gets the expressions for
$\omega_{21}$, $\nu_{22}$ and $\nu_{21}$ by interchanging
$1\leftrightarrow 2$.

In the case of a three-dimensional system ($d=3$), all the above
results reduce to those previously obtained for hard spheres when
one takes Maxwellian distributions for the reference homogeneous
cooling state. \cite{GD02,MGD06}


\begin{thebibliography} {99}

\bibitem{GS95}A. Goldshtein and M. Shapiro, Mechanics of collisional motion of granular materials.
Part 1. General hydrodynamic equations, {\em J. Fluid Mech.} {\bf
282}:75--114 (1995).

\bibitem{BDS97}J. J. Brey, J. W. Dufty, and A. Santos, Dissipative
dynamics for hard spheres, {\em J. Stat. Phys.} {\bf
87}:1051--1066 (1997).

\bibitem{BP04}N. V. Brilliantov and T. P\"oschel, {\em Kinetic Theory
of Granular Gases} (Oxford University Press, Oxford, 2004).

\bibitem{CC70}S. Chapman and T. G. Cowling, {\em The Mathematical Theory of Nonuniform Gases}
(Cambridge University Press, Cambridge, 1970).

\bibitem{simple}J. J. Brey, J. W. Dufty, C. S. Kim, and A. Santos,
Hydrodynamics for granular flow at low-density, {\em Phys. Rev. E}
{\bf 58}:4638--4653 (1998); J. J. Brey and D. Cubero, Hydrodynamic
transport coefficients of granular gases, in {\em Granular Gases},
edited by T. P\"oschel and S. Luding, Lecture Notes in Physics
(Springer Verlag, Berlin, 2001), pp. 59--78; V. Garz\'o and J. M.
Montanero, Transport coefficients of a heated granular gas, {\em
Physica A} {\bf 313}:336--356 (2002).

\bibitem{mixture}J. T. Jenkins and F. Mancini, Kinetic theory for binary
mixtures of smooth, nearly elastic spheres, {\em Phys. Fluids A}
{\bf 1}:2050--2057 (1989); P. Zamankhan, Kinetic theory for
multicomponent dense mixtures of slightly inelastic spherical
particles, {\em Phys. Rev. E} {\bf 52}:4877--4891 (1995); B.
Arnarson and J. T. Willits, Thermal diffusion in binary mixtures
of smooth, nearly elastic spheres with and without gravity, {\em
Phys. Fluids} {\bf 10}:1324--1328 (1998); J. T. Willits and B.
Arnarson, Kinetic theory of a binary mixture of nearly elastic
disks, {\em Phys. Fluids} {\bf 11}:3116--3122 (1999).



\bibitem{GD99b}  V. Garz\'{o} and J. W. Dufty, Homogeneous cooling
state for a granular mixture, {\em Phys. Rev. E} {\bf
60}:5706--5713 (1999).

\bibitem{MP99}P. A. Martin and J. Piasecki, Thermalization of a particle by
dissipative collisions, {\em Europhys. Lett.} {\bf 46}:613--616
(1999).

\bibitem{computer}See for instance, J. M. Montanero and V. Garz\'o,
Monte Carlo simulation of the homogeneous cooling state for a granular mixture, {\em Gran. Matt.} {\bf 4}:17--24
(2002); A. Barrat and E. Trizac, Lack of energy equipartition in homogeneous heated binary granular mixtures,
{\em ibid.} {\bf 4}:57--63 (2002); S. R. Dahl, C. M. Hrenya, V. Garz\'o, and J. W. Dufty, Kinetic temperatures
for a granular mixture, {\em Phys. Rev. E} {\bf 66}:041301 (2002); R. Pagnani, U. M. B. Marconi, and A. Puglisi,
Driven low density granular mixtures, {\em ibid.} {\bf 66}:051304 (2002); D. Paolotti, C. Cattuto, U. M. B.
Marconi, and A. Puglisi, Dynamical properties of vibrofluidized granular mixtures, {\em Gran. Matt.} {\bf
5}:75--83 (2003); P. Krouskop and J. Talbot, Mass and size effects in three-dimensional vibrofluidized granular
mixtures, {\em Phys. Rev. E} {\bf 68}: 021304 (2003); H. Wang, G. Jin, and Y. Ma, Simulation study on kinetic
temperatures of vibrated binary granular mixtures, {\em ibid.} {\bf 68}:031301 (2003); J. J. Brey, M. J.
Ruiz-Montero, and F. Moreno, Energy partition and segregation for an intruder in a vibrated granular system
under gravity, {\em Phys. Rev. Lett.} {\bf 95}:098001 (2005); M. Schr\"oter, S. Ulrich , J. Kreft , J. B. Swift,
and H. L. Swinney, Mechanisms in the size segregation of a binary granular mixture, {\em Phys. Rev. E} {\bf
74}:011307 (2006).


\bibitem{exp1}R. D. Wildman and D. J. Parker, Coexistence of two granular
temperatures in binary vibrofluidized beds, {\em Phys. Rev. Lett.} {\bf 88}:064301 (2002); K. Feitosa and N.
Menon, Breakdown of energy equipartition in a 2D binary vibrated granular gas, {\em ibid.} {\bf 88}:198301
(2002).

\bibitem{JM87}J. Jenkins and F. Mancini, Balance laws and constitutive relations
for plane flows of a dense binary mixture of smooth, nearly elastic, circular disks, {\em J. Appl. Mech.} {\bf
54}:27--34 (1987).


\bibitem{GD02}V. Garz\'o and J. W. Dufty, Hydrodynamics for a granular binary mixture
at low density, {\em Phys. Fluids} {\bf 14}:1476--1490 (2002).

\bibitem{GA05}V. Garz\'o  and A. Astillero, Transport coefficients
for inelastic Maxwell mixtures, {\em J. Stat. Phys.} {\bf
118}:935--971 (2005).



\bibitem{MGD06}V. Garz\'o, J. M. Montanero, and J. W. Dufty, Mass
and heat fluxes for a binary granular mixture at low-density, {\em
Phys. Fluids} {\bf 18}:083305 (2006).

\bibitem{GM04}V. Garz\'o and J. M. Montanero, Diffusion of impurities in a granular gas,
{\em Phys. Rev. E} {\bf 69}: 021301 (2004).

\bibitem{MG03}J. M. Montanero and V. Garz\'o, Shear viscosity for a heated granular binary mixture
at low-density, {\em Phys. Rev. E} {\bf 67}:021308 (2003).

\bibitem{SGNT06}D. Serero, I. Goldhirsch, S. H. Noskowicz, and M.-L. Tan,
Hydrodynamics of granular gases and granular gas mixtures, {\em J. Fluid Mech.} {\bf 554}:237--258 (2006).

\bibitem{exp}See for instance,
S. Warr, J. M. Huntley, and G. T. H. Jacques, Fluidization of a
two-dimensional granular system: Experimental study and scaling
behavior, {\em Phys. Rev. E} {\bf 52}:5583--5595 (1995); J. S.
Olafsen and J. S. Urbach, Velocity distribution and density
fluctuations in a granular gas, {\em ibid.} {\bf 60}:R2468--R2471
(1999); R. D. Wildman, J. M. Huntley, and J. P. Hansen,
Self-diffusion of grains in two-dimensional vibrofluidized bed,
{\em ibid.} {\bf 60}:7066--7075 (1999); F. Rouyer and N. Menon,
Velocity fluctuations in a homogeneous 2D granular gas in steady
state, {\em Phys. Rev. Lett.} {\bf 85}:3676--3679 (2000); D. L.
Blair and A. Kudrolli, Velocity correlations in dense granular
flows, {\em Phys. Rev. E} {\bf 64}:050301 (2001); S. Horluck, M.
van Hecke, and P. Dimon, Shock waves in two-dimensional granular
flow: Effects of rough walls and polydispersity, {\em ibid.} {\bf
67}:021304 (2003).

\bibitem{Bird}G. A. Bird, {\em Molecular Gas Dynamics and the Direct
Simulation Monte Carlo of Gas Flows} (Clarendon, Oxford, 1994).

\bibitem{SGD04}A. Santos, V. Garz\'o, and J. W. Dufty, Inherent
rheology of a granular fluid in uniform shear flow, {\em Phys.
Rev. E} {\bf 69}:061303 (2004).

\bibitem{NE98}T. P. C. van Noije and M. H. Ernst, Velocity
distributions in inhomogeneous granular fluids: the free and the heated case,
{\em Gran. Matt.} {\bf 1}:57--64(1998).

\bibitem{G02}V. Garz\'o, Tracer diffusion in granular shear flows, {\em Phys. Rev.
E} {\bf 66}:021308 (2002).


\bibitem{BRM05}J. J. Brey, M. J. Ruiz-Montero, and F. Moreno, Energy partition
and segregation for an intruder in a vibrated granular system
under gravity, {\em Phys. Rev. Lett.} {\bf 95}:098001 (2005);
Hydrodynamic profiles for an impurity in an open vibrated granular
gas, {\em Phys. Rev. E} {\bf 73}:031301 (2006).

\bibitem{BC01}J. J. Brey and D. Cubero, Hydrodynamic
transport coefficients of granular gases, in {\em Granular Gases},
edited by T. P\"oschel and S. Luding, Lecture Notes in Physics
(Springer Verlag, Berlin, 2001), pp. 59--78.


\bibitem{M89}J. A. McLennan, {\em Introduction to Nonequilibrium Statistical Mechanics}
(Prentice-Hall, New Jersey, 1989).


\bibitem{BRCG00} J. J. Brey, M. J. Ruiz-Montero, D. Cubero, and R. Garc\'{\i}a-Rojo, Self-diffusion in
freely evolving granular gases, {\em Phys. Fluids} {\bf
12}:876--883 (2000).


\bibitem{GM03}V. Garz\'o and J. M. Montanero, Shear viscosity for a moderately dense granular
binary mixture, {\em Phys. Rev. E} {\bf 68}:041302 (2003).


\bibitem{MSG05}J. M. Montanero, A. Santos, and V. Garz\'o, DSMC evaluation of the Navier-Stokes
shear viscosity of a granular fluid, in {\em Rarefied Gas Dynamics
24}, edited by M. Capitelli (American Institute of Physics, Vol.
762, 2005), pp. 797--802; preprint arXiv: cond-mat/0411219.

\bibitem{BRM04}J. J. Brey and M. J. Ruiz-Montero, Simulation
study of the Green-Kubo relations for dilute granular gases, {\em
Phys. Rev. E} {\bf 70}:051301 (2004).

\bibitem{BRMMG05}J. J. Brey, M. J. Ruiz-Montero, P. Maynar, and M.
I. Garc\'{\i}a de Soria, Hydrodynamic modes, Green-Kubo relations,
and velocity correlations in dilute granular gases, {\em J. Phys.:
Condens. Matter} {\bf 17}:S2489--S2502 (2005).

\bibitem{MSG06}J. M. Montanero,  A. Santos, and V. Garz\'o, First-order
Chapman-Enskog velocity distribution function in a granular gas,
{\em Physica A} {\bf 376}:75--93 (2007).

\bibitem{GSM06}V. Garz\'o, A. Santos, and J. M. Montanero,
Modified Sonine approximation for the Navier-Stokes transport
coefficients of a granular gas, {\em Physica A} {\bf 376}:94--107 (2007).

\bibitem{MCK83}M. L\'opez de Haro, E. G. D. Cohen, and J. M.
Kincaid, The Enskog theory for multicomponent mixtures. I. Linear
transport theory, {\em J. Chem. Phys.} {\bf 78}:2746--2759 (1983).

\bibitem{GD99a}V. Garz\'{o} and J. W. Dufty, Dense fluid transport for inelastic hard spheres,
{\em Phys. Rev. E} {\bf 59}:5895--5911 (1999).


\bibitem{EB02}M. H. Ernst and R. Brito, Scaling solutions of inelastic Boltzmann
equations with over-populated high energy tails, {\em J. Stat.
Phys.} {\bf 109}:407--432 (2002).

\bibitem{note}Some misprints occur in the expressions given in
Ref.\ \onlinecite{GD02} for the heat flux. The results displayed
here correct and extend such results to $d$ dimensions.


\end{thebibliography}
\end{document}